\documentstyle[12pt,epsf]{article}
\def\beq{\begin{equation}}   \def\eeq{\end{equation}}

\def\lsim{\mathrel{\rlap{\lower3pt\hbox{\hskip0pt$\sim$}}
    \raise1pt\hbox{$<$}}}         
\def\gsim{\mathrel{\rlap{\lower4pt\hbox{\hskip1pt$\sim$}}
    \raise1pt\hbox{$>$}}}         

\begin{document}
\begin{titlepage}

\begin{flushright}
TPI-MINN-00/44\\
UMN-TH-1920/00\\
hep-th/0009131
\end{flushright}

\vspace{0.3cm}

\begin{center}
\baselineskip25pt

{\Large\bf
Quark-Hadron Duality$\,$\footnote{Based on the talks
delivered at the VIII-th International Symposium on Heavy Flavor Physics,
Southampton, UK, 25--29 July 1999, and the International Workshop
{\em Gribov--70},   Orsay, France, 27--29 March  2000. To be published in
the Boris Ioffe Festschrift {\em At the 
Frontier of Particle Physics/Handbook of
QCD}, Ed. M. Shifman (World Scientific, Singapore, 2001).}}

\end{center}

\vspace{0.3cm}

\begin{center}

{\large  M. Shifman}

\vspace{0.2cm}
{\em Theoretical Physics Institute, University of Minnesota,
Minneapolis,
MN 55455}

\vspace{1cm}

{\large\bf Abstract}

\vspace*{.25cm}

\end{center}

I review the notion of the quark--hadron duality 
from the modern perspective.  Both, the theoretical foundation and
practical applications are discussed. The proper theoretical framework in which
the problem can be formulated and treated is Wilson's operator product
expansion (OPE).  Two models developed for the description of  duality
violations are considered in some detail: one is 
 instanton-based, another resonance-based. The mechanisms they 
represent are
complementary. Although both models are rather 
primitive
(their largest virtue is their simplicity) they hopefully capture  
 important features of the phenomenon. Being open for improvements,
they can be used ``as is" for orientation 
in the studies of duality violations in the
processes of practical interest.

\end{titlepage}
\section{Introduction}

Quantum chromodynamics (QCD) is a very strange theory.
All theoretical  calculations are done in terms of quarks and gluons.
At the same time,  quarks and gluons are never detected experimentally.
What is actually produced and detected in experimental devices are
hadrons: pions, kaons, protons, {\em etc.} The quark-hadron duality allows one,
under certain circumstances, to bridge the gap
between the theoretical predictions and experimentally observable quantities.
The idea was first formulated at the dawn of the QCD era
by Poggio, Quinn and Weinberg \cite{Poggio:1976af},
who suggested that certain inclusive  hadronic cross
sections at high energies, being appropriately averaged over an  energy range,
had to  (approximately) coincide
with the cross sections one could calculate in
the quark-gluon  perturbation theory.
The uses of this theoretical construction are countless:
the $e^+e^-$ annihilation, deep inelastic scattering, hadronic $\tau$ decays,
inclusive decays of heavy quarks, physics at the $Z$ peak, to name just a few.
In spite of the dramatic developments in QCD in the subsequent two decades
the notion of the quark-hadron duality remained vague, essentially at the
1976 level, with the very basic questions unanswered. These questions are:

\vspace{01.cm}

$\bullet$  What energy is considered to be high enough for the quark-hadron
duality to set in, and what accuracy is to be expected?

$\bullet$ What weight function is appropriate for the averaging of the
experimental cross sections?

$\bullet$ If the theoretical prediction includes only perturbation theory,
should one limit oneself to some particular
order in the  the $\alpha_s$ series?

$\bullet$ Do we have to include known nonperturbative effects
(e.g. condensates) in the theoretical prediction?

$\bullet$ Given a definition of the quark-hadron duality,
can one estimate deviations from duality and how?

\vspace{01.cm}

Of course, not all of these questions are independent.
For instance, answering the last question, one will simultaneously learn
the boundary  energy.

Systematic explorations of these and related issues started in earnest
 in 1994
\cite{Shifman:1994yf}.  As in many other instances, this was dictated by
practical needs. Previously, the accuracy of the experimental data
on hard inclusive processes was rather modest, so that the 
Poggio-Quinn-Weinberg prescription was good enough. 
By 1994 the data, mostly associated with the $b$ quark physics, 
became so precise and the questions  raised so acute, that
a much better theoretical understanding became imperative. 
Probably,
the most clear-cut example is the problem of the $B$ semileptonic branching
ratio \cite{Bigi:1994fm}: theoretical expectations 
obtained in the quark-gluon language exceed the measured number
by 10 to 20\%.  Possible deviation from duality is suspected to be
a major source
of  theoretical uncertainties. If one could reliably rule out 
duality violations at
this level, the conclusion of the interference of new physics would 
ensue (provided the experimental data stay intact, of course).
The stakes are
quite high.

It is fair to say that (short of the full solution of QCD)
understanding and controlling the accuracy of the quark-hadron
duality is one of the most important and challenging problems
for the QCD practitioners today.
In this issue one cannot expect help from lattices.
The duality violation is a phenomenon inseparable from
the Minkowskian kinematics; numerical Euclidean approaches,
such as lattice QCD, have nothing to say on this issue.
Analytic methods are needed.

 In this review I will summarize the results
of an  investigation
\cite{Shifman:1995mt}--\cite{Lebed:2000gm},
which spans over six years and is not yet complete.
I will discuss the modern formulation of the problem
 based on  Wilson's operator product
expansion (OPE). This formulation is solid and unambiguous.
Then I will review models which were designed to give us 
a certain idea
of (and a degree of control over) the
deviations from duality.  There are two classes of such models: instanton-based
and resonance-based. They are complementary to each other; I will discuss both
classes. Finally, I will present sample applications
in the processes of the current interest, such as the hadronic $\tau$ decays.

\section{The Quark-Hadron Duality: What Does That Mean?}

Let us consider an idealized theory -- QCD, with  two massless quarks,
$u$ and $d$. We are interested in the total hadronic cross section of
the $e^+ e^-$  annihilation. The ``photon" in our theory is idealized too,
its coupling to the quark current has the form
\begin{equation}
J_\mu = \frac{\bar u\gamma_\mu u -\bar d\gamma_\mu 
d}{\sqrt{2}}\, .
\label{currents}
\end{equation}
(In fact, this is the isovector part of the actual electromagnetic current).
We define the two-point function $\Pi_{\mu\nu}$  
\beq
\Pi_{\mu\nu} = i \int {\rm e}^{iqx} d^4 x \langle 0|T\{ J_\mu (x) 
J^\dagger_\nu 
(0) \} |0\rangle\, .
\label{Tproduct}
\eeq
Here $q$ is the total momentum of the quark-antiquark pair.
Due to the current conservation $\Pi_{\mu\nu}$ is transversal,
\beq
 \Pi_{\mu\nu}
= (q_\mu q_\nu -q^2g_{\mu\nu})\Pi 
(q^2)\, .
\label{curco}
\eeq
The experimentally observable quantity is the imaginary part 
of $\Pi (q^2)$ at positive values
of $q^2$ (i.e. above the physical threshold of the hadron production),
the spectral density,
\beq
\rho (s) =\frac{12\pi}{N_c}\, \mbox{Im}\,\Pi (s)\, , \,\,\,  s\equiv 
q^2\, .
\label{rhoim}
\eeq
Up to a normalization, 
the expression above coincides with the cross section of $e^+e^-$ annihilation into 
hadrons measured in the units $\sigma (e^+e^-\rightarrow
\mu^+\mu^-)$, the famous ratio $R$. 

Theoretically one can calculate $\Pi (q^2)$ in the deep Euclidean domain,
at negative $q^2$. For instance, from the free quark 
loop of Fig. 1 one gets
\beq
\Pi (Q^2) \rightarrow -\frac{N_c}{12\pi^2}\ln Q^2\, \quad Q^2 \equiv -q^2\,.
\label{lqpi}
\eeq
Performing the analytic continuation to the Minkowski domain
and taking the imaginary part one arrives at
\beq
\rho (s)_{\rm theor}\to 1\, ,\quad  s\to\infty\,.
\label{lqpidop}
\eeq

\begin{figure}   
\epsfxsize=6cm
\centerline{\epsfbox{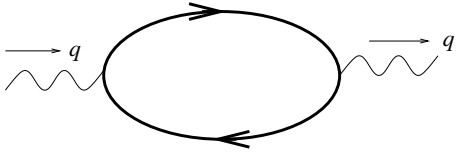}}
 \caption{The one-loop graph determining
the polarization operator and the spectral density in the leading
(parton) approximation. The ``photon" momentum is denoted by $q$.}
\end{figure}

This is the spectral density in the theory with free quarks, i.e.
$\alpha_s =0$.
There are various corrections to the free quark result
(\ref{lqpi}). The perturbative gluon
exchanges give rise to the $\alpha_s (Q^2)$ series. Nonperturbative
(power) corrections come from the quark and gluon condensates
and from other sources, e.g. the small-size instantons.
A systematic method of handling the theoretical calculations of
$\Pi (Q^2) $ in the {\em deep Euclidean} domain is provided
by Wilson's operator product expansion (OPE)
\cite{Wilson}. Essentially, this is a bookkeeping procedure: one consistently
separates the short-distance contributions  (i.e. those coming from distances
$\leq \mu^{-1}$) from the large distance contributions
(i.e. those coming from distances
$\geq \mu^{-1}$). Here $\mu$ is a theoretical parameter
(usually referred to as the normalization point) separating the two domains.
The choice of $\mu$ is a matter of convenience --  observable quantities
do not depend on it. 

The short-distance contributions determine the coefficients $C_n(q)$ of OPE ,
\begin{equation}
D(q^2)  =\sum_n C_n(q;\mu )\langle {\cal O}_n(\mu )\rangle \,,\qquad
D(Q^2) \equiv - (4\pi^2) Q^2 (d\Pi / dQ^2 )\, .
\label{tee}
\end{equation}
The normalization point $\mu$ is indicated explicitly.  
 The sum in Eq. (\ref{tee}) runs over all possible 
Lorentz and gauge invariant local operators built from the gluon and 
quark fields.  The operator  of  the lowest (zero) dimension is 
the 
unit 
operator {\bf I},
followed by the gluon condensate $G_{\mu\nu}^2$, of dimension 
four.
The four-quark condensate gives an example of dimension-six
operators.

At short distances QCD is well described by 
the quark-gluon perturbation theory.
 Therefore,
as a first approximation, it is reasonable to  calculate 
$C_n$ perturbatively, in the form of expansion in $\alpha_s (Q) \sim
(\ln Q)^{-1}$. This certainly does not mean that the coefficients  
$C_n$ are free from nonperturbative (nonlogarithmic) terms.
The latter may and do appear in $C_n$'s; they are of the type
$\sim Q^{-\gamma}$ where $\gamma$ is a positive number, not necessarily an
integer.  Such terms in $C_n$'s are generated, for instance, by the small-size
instantons.  Another source of the power terms in $C_n$,
of a technical rather than dynamical nature, is  the
normalization point $\mu$: in calculating $C_n (\mu )$
one must remove all soft contributions with off-shellness less than $\mu$.

The condensate terms in Eq. (\ref{tee}) give rise to corrections
of  the type $(\Lambda_{\rm QCD}/Q)^{n}$ where $n$ is an integer $\geq 4$,
the (normal) dimension of the operator $ {\cal O}_n$, 
modulo logarithms associated with the anomalous dimensions.
$\Lambda_{\rm QCD}$ is the scale parameter of QCD
(sometimes, I will drop the subscript ``QCD" for brevity)
entering through the matrix elements of $ {\cal O}_n$'s.

If one could calculate $\Pi (Q^2)$ in the Euclidean domain {\em exactly},
one could analytically continue the result to the Minkowski domain,
and then take the imaginary part.  The spectral density
$\rho (s)_{\rm theor}$ obtained in this way would present the {\em  exact}
theoretical prediction for the measurable hadronic  cross section.
{\em There would be no need for duality.}

In practice, our calculation of $\Pi (Q^2)$ is approximate, for many reasons.
First, nobody is able to calculate the infinite
$\alpha_s (Q^2)$ series for the coefficient functions, let alone the infinite
condensate series. Both have to be truncated at some finite order.
A few lowest-dimension condensates that can be captured,
are known approximately.  The best we can do is
analytically continue the {\em truncated} theoretical 
expression, term by term,
from positive to negative $Q^2$. For each term in the 
expansion the imaginary
part at  positive $q^2$ (negative $Q^2$) is well-defined.  We assemble 
them together and declare
the corresponding $\rho (s)_{\rm theor}$ to be dual to
the hadronic cross section $\rho (s)_{\rm exp}$.
In the given context ``dual" means  equal.

Let me elucidate this point in more detail.
Assume that $\Pi (Q^2)$ is calculated through $\alpha_s^2$
and $1/Q^4$, while the 
terms  $\alpha_s^3$ and $1/Q^6$ (with possible logarithms)
are dropped.
Then the theoretical quark-gluon 
spectral density, obtained as described above, is expected to 
coincide with $\rho
(s)_{\rm exp}$, with the uncertainty of order $O[(\alpha_s (s))^3]$
and $O(1/s^3)$.
The uncertainty 
in the theoretical prediction of this order of magnitude is {\em natural} since
 terms of this order are neglected in
the theoretical calculation of  $\Pi (Q^2)$. If the coincidence
in this corridor does take place,
we say that the quark-gluon prediction is dual to the hadronic
spectral density. If there are deviations going {\em beyond the 
natural uncertainty},
we call them violations of duality. 
Needless to say that, once our calculation of 
$\Pi (Q^2)$ becomes more precise, the definition of the
``natural uncertainty" in $\rho (s)_{\rm theor}$ changes accordingly.

This is the most clear-cut definition  
I can suggest. From the formal standpoint, it connects
the duality violation issue with that
of analytic continuation from the Euclidean to Minkowski domain.
Negligibly small corrections (legitimately) omitted in the
Euclidean calculations may and do get enhanced in Minkowski.

 Before I proceed to 
explain, from the physical standpoint,  where
the violations of duality come from, and their pattern,
I would like to make a  remark
important in the conceptual  aspect.
The necessity of truncation of the $\alpha_s$ and 
condensate series is not just due to our practical inability to calculate 
high-order terms. (For the vast majority of
the  theorists, myself including,
``high-order" begins at the next-to-next-to-leading level.)
Assume  we have a ``sorcerer's stone" which would allow us
to exactly calculate any term in the expansion we want. Still, 
we would  not be able to 
find $\Pi (Q^2)_{\rm exact}$ because the both series are factorially divergent.
The $\alpha_s$ series has at least two known sources of the
factorial behavior: first, the number of the Feynman  graphs grows
factorially at high orders \cite{hofd}; second, there are graphs with 
renormalons,
(of which we will not bother 
in what follows since the infrared  renormalons 
-- the only ones which are potentially
dangerous -- are 
totally eliminated by the introduction of the
normalization point $\mu$ \cite{Mueller:1992xz}).
The condensate series is divergent too. 
The factorial
divergence of the condensate series is studied even to a lesser extent
than that of the $\alpha_s$ series. 
The only fact considered to be firmly established
\cite{Shifman:1994yf,Shifman:1995mt} is the divergence
{\em per se}. It is not Borel-summable, which sets
a limit of the theoretical accuracy.
Including more and more terms in the series
would not help, and even an optimal truncation
would leave a gap between $\Pi (Q^2)_{\rm theor}$
and $\Pi (Q^2)_{\rm exact}$.

\section{Where the Duality Violations Come From?}

Theoretical  calculations of 
inclusive processes in QCD are performed through 
Wilson's OPE in the Euclidean domain.
Therefore, to understand what is included in such
calculations and what is left out, one should have a clear picture
of  limitations of OPE.

Let us return to the consideration of  the correlation function of two
currents,  $$\langle 0|T\{ J_\mu (x) 
J^\dagger_\nu 
(0) \} |0\rangle \,,$$ which determines the polarization operator in Eq.
(\ref{Tproduct}), at negative (Euclidean) $x^2$.
Wilson's OPE is nothing but an expansion of $\langle 0|T\{ J_\mu (x) 
J^\dagger_\nu 
(0) \} |0\rangle $ in {\em singularities} at $x^2=0$.
It  properly captures all terms of the type
$1/x^6$ times logarithms, or $\ln x^2$, or $x^2\ln x^2$,
and so on. Every term of this type is represented 
in the condensate series provided that the calculation is 
carried out to a sufficient order. 

Let me note in passing that limiting oneself to  singular terms 
(and, in particular,
to the leading singular term) is a reasonable approximation in the theories
in which the asymptotic conformal regime sets in at short distances in
the power-like manner. In fact, OPE in its original formulation was designed by
Wilson for applications in such
theories (see the first work in Ref. \cite{Wilson}). In QCD 
the approach to the asymptotically free regime is very slow -- logarithmic.
 That's why the precision of the
leading (parton) approximation is sufficient only for
very rough estimates, and that's why the 
high-order terms, both logarithmic and
power,  must be kept. The time when the QCD practitioners would
be satisfied by the leading, or even the first
subleading approximation, is long gone. 

It is clear  that the function
$\langle 0|T\{ J_\mu (x) 
J^\dagger_\nu 
(0) \} |0\rangle $ is not fully determined by its singularities
at $x^2=0$. Generally speaking, one may have,
additionally, isolated  singularities at
finite $x^2$, or a singularity
at infinity, which are not reflected in the truncated
Wilson's OPE.  Consider, say, a singularity at finite $x^2$  of the form
\beq
\frac{1}{x^2+\rho^2}\, . 
\label{ssin}
\eeq
The expansion of this function (truncated at any order)
generates  derivatives of $\delta (q^2)$ in the momentum space;
therefore, it has no impact on the OPE expansion at large $Q^2$.
This contribution is clearly missing. The Fourier transform of
Eq. (\ref{ssin}) at large Euclidean $Q^2$ falls off 
as $\exp (-Q\rho )$. Thus, this term is  smaller than any of the terms in
the condensate expansion. We should not forget, however,
that our final goal is a prediction in the Minkowski domain.
Upon  analytic continuation,
 the exponentially small  term $\exp (-Q\rho )$ looses its suppression 
and becomes
 oscillating, $\sin (-E\rho )$, where $E$ stands for the total energy,
$E=\sqrt{q^2}$. 
If it were not
for the power suppression in the pre-exponent, 
such exponential/oscillating terms
would be a total disaster. Since they are not seen in OPE, any prediction
for the inclusive cross sections  made
through OPE (which is equivalent, in practice, to perturbation theory plus
a few 
condensates) would be grossly wrong. There would be no quark-hadron
duality at all. All calculations of the hard
processes -- from the total hadronic cross sections  in  $e^+e^-$ annihilation,
to jet physics, to the heavy quark decays -- would be valid
roughly  up to a factor of
two,  no matter how large the energy release is.

Fortunately, one avoids the disaster due to the fact that
 the duality-violating exponential/oscillating terms
are, in fact, suppressed as $E^{-\kappa}\, \sin (-E\rho )$ where 
$\kappa$ is a positive index, which depends on the process
under consideration and is typically rather large. This justifies
theoretical predictions based on a few first terms in the 
condensate expansion. Needless to say that determining
$\kappa$ 
is one of the most important tasks
in the issue of the quark-hadron duality.

Singularities at $x^2\to\infty$
also  lead to the exponential/oscillating terms
of a somewhat different form, of the type $\exp (-Q^2\rho^2 )$ in Euclidean.
In the Minkowski domain one gets an oscillating function of the
type $E^{-\eta}\, \sin (-E^2\rho^2 )$, where $\eta$
is another positive index. In one and the same process, one can expect
both duality-violating components to show up.
Generally speaking, the indices $\kappa$ and $\eta$ are unrelated;
at least, at the moment we do not see any obvious relation between them.
(The coincidence of  $\kappa$ and $\eta$ in the  practically important problem
of the inclusive hadronic $\tau$ decays, $\kappa= \eta = 6$,
seems to be accidental,  see Sec. 8).

How can one isolate the singularities at $x^2 =0$
from others 
diagrammatically?
To answer this question it is convenient
to pass to the momentum space. In the leading approximation
the polarization operator $\Pi (Q^2)$
is presented  by the graph of Fig. 1.
The large Euclidean momentum $q$ (remember, $Q^2\to\infty
$) flows in the photon line
on the left, propagates through the fermion lines, and leaves
the graph  through the photon line
on the right. The virtual momenta of the fermion lines typically scale as $q$.
The first $\alpha_s$ correction is presented in Fig. 2.
In this graph the virtual momenta of all  lines in the loops,
the fermion and the gluon, scale as $q$. Thus, the diagram of Fig. 2
determines the leading logarithmic correction
to the parton result (\ref{lqpi}) or (\ref{lqpidop}).

\begin{figure}   
\epsfxsize=6cm
\centerline{\epsfbox{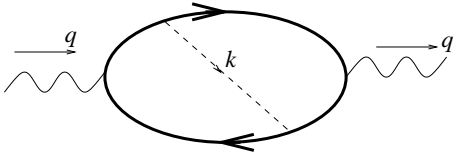}}
 \caption{The one-gluon correction in the
the polarization operator. The gluon momentum is denoted by $k$.}
\end{figure}

Although the leading contribution to the integral comes 
from the domain where
all virtual momenta are proportional to $q$,
there are nonvanishing contributions from other kinematical domains.
For instance, one can consider a ``corner"
where the gluon virtual momenta $k$  in Fig. 2 is small,
$|k| <\mu $, and does not scale with  $q$. 
Certainly, at small $k$ the gluon propagator is {\em not} given by
perturbation theory. The gluon line must be cut.
This corresponds to the gluon
condensate term \cite{SVZ} in the condensate expansion. 

In general, any term in the
condensate expansion can be interpreted in this way, in terms of
{\em factorization in the momentum space}. The large external momentum
$q$ is transmitted through one or several hard lines -- their
 momenta scale  with $q$. This is the definition of
``hardness." The remainder of the graph is
a soft part, which  factors out and gives rise to  gluon, quark and mixed
condensates. The virtual momenta in this part of the graph
do not scale with $q$, they are limited by a fixed parameter $\mu$
(this is the definition of
``softness"). 

The hard part of the graph is responsible for the coefficient
functions. The fact that not all lines in the given graph are hard
results in the power suppression of the corresponding coefficient function.
Letting more and more lines to be soft, one obtains
consecutive terms in the condensate expansion. What is missing?

It is conceivable that the number of lines through which $q$ is 
transmitted becomes so large, that though $Q\to\infty$, neither of the lines is
hard. Of course, in this case one cannot speak of the
separate lines in the Feynman graphs. It would be more relevant to say
that the external momentum $q$ is transmitted from the incoming to the
outgoing photon through a coherent soft field fluctuation, see Fig. 3.
An example is provided, for instance, by a fixed-size instanton.
\begin{figure}   
\epsfxsize=8cm
\centerline{\epsfbox{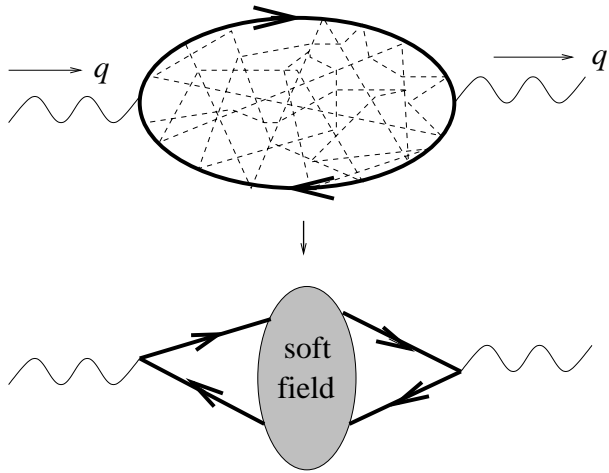}}
 \caption{Transmitting a large external momentum through a soft field.}
\end{figure}
It is clear that this mechanism is conceptually related to
the truncated tail of the condensate series. Indeed,
as one proceeds to higher condensates, more lines become soft.
Eventually we arrive at the situation when all lines are soft.
Mathematically, the exponential/oscillating contribution
is related to the factorial divergence of the condensate
series \cite{Shifman:1995mt}. This is the usual story.
The exponentially small terms in  Euclidean convert into an
oscillating function in Minkowski.

I will use the fixed-size instantons
for the purpose of modeling this mechanism of duality violations.
The observation that ``soft" instantons
generate an oscillating component ascends to Refs.
\cite{DubSm,Maggiore}.
By no means I imply that the instantons
are the dominant soft fields in the QCD vacuum. True, there are models 
in which they are assumed to be dominant,
 the so-called instanton liquid models \cite{Shur}.
Their status in the range of questions I am interested in
(the duality violations) has yet to be clarified.  
I will use instantons for the purpose of orientation, in  
 hope that at least some  features of the
results obtained in this way will be more general than
 the model itself.

\section{Model or Theory?}

The topic I address -- the quark-hadron duality violations --
has a unique status.
By definition,  one 
 can{\em not} build  an exhaustive {\em theory} of the duality
violations based   on the
first principles. Indeed, assuming there is a certain dynamical mechanism
(which goes beyond perturbation theory and condensates)
for which such a theory exists, one will immediately include
the corresponding component in the theoretical calculation.
The reference quantity, $\Pi (Q^2)_{\rm theor}$, will be redefined
accordingly. After the analytic continuation to Minkowski,
this will lead, in turn,  to a new
theoretical spectral density to be used as a reference $\rho (s)_{\rm theor}$
in the duality relation. 

Thus, by the very nature of the problem, it is bound to be treated in models
of various degrees of fundamentality and reliability.
This is because the duality violation  parametrizes our ignorance.
Ideally, the models one should aim at
must have a clear physical interpretation,
and must be tested,  in their key features, against experimental data.
This will guarantee a certain degree of confidence
when these models are  applied to the estimates
of the duality violations
in the processes and kinematical conditions where they had not
been  tested.

\section{The Physical Picture Behind the Duality}

Before delving into technical details
I will describe the  phenomenon from a slightly different perspective.
The quark-hadron duality takes place
in those  processes where one can isolate two stages in the process under
consideration, occurring at two distinct scales. A basic transition
involving quarks (gluons) must typically  occur at 
a short scale regulated by external parameters
such as $Q$, $m_Q$, etc. For instance, in the $e^+e^-$
annihilation the basic transition is the conversion of the 
virtual $\gamma$ into $\bar u u$ or $\bar d d$.
Then, at the second stage,  the quarks (gluons) materialize
in the form of hadrons, at a much larger scale. 
In the appropriate frame, the first time scale 
is of order $1/Q$ while the second of order $Q/\Lambda^2$. 
By that time, the original quarks are far away from each other
-- a residual interaction cannot significantly alter the transition  cross 
section which was ``decided" at the first (quark-gluon) stage.

The duality violations are  due to (i) rare  atypical events,
when the basic quark transition occurs at large rather than short distances;
(ii) residual interactions occurring at large distances
between the quarks produced at short distances. In the first case
appropriate (Euclidean) correlation functions
develop singularities at finite $x^2$, while the second
mechanism is correlated with the $x^2\to\infty$ behavior.

In both cases the duality violating component follows the pattern
I have discussed above -- {\em exponential  in Euclidean
and oscillating  in Minkowski}.  Three distinct regimes were
identified and considered in the literature so far:

\vspace{0.3cm}

$\bullet$ (i) Finite-distance singularities
\beq
s^{-\kappa /2}\, \sin (\sqrt{s} )\,;
\label{SS}
\eeq

\vspace{0.1cm}

$\bullet$ (ii) Infinite-distance singularities ($N_c = \infty$)
\beq
s^{-\eta /2}\, \sin (s)\,;
\label{SSS}
\eeq

\vspace{0.1cm}

$\bullet$ (iii) Infinite-distance singularities ($N_c $ large but finite, $s\to\infty$)
\beq
\exp{(-\alpha s)} \sin (s), \quad \alpha =
O\left(\frac{1}{N_c}\right) \ll 1\, .
\label{SSSS}
\eeq

\vspace{0.1cm}

These regimes are not mutually exclusive -- in concrete processes
one may expect
the duality violating component to be a combination of
 (i) and (ii), or (i) and (iii).
From the theoretical standpoint it is quite difficult
to consistently define the duality violating component of the type (3).
An operational definition I might suggest is as follows:
Start from the limit $N_c = \infty$ and identify
the component of the type (2).  Follow its evolution as 
$N_c$ becomes large but finite.

\section{An Instanton-Based Model}

The basic features of the formalism \cite{Chibisov:1997wf} to be used below
to model exponential/oscil\-lating terms (which present
deviations from duality) are as follows:

\vspace{0.1cm}

(i) One considers  quarks propagating in the
instanton background field.  The instanton size $\rho$ is assumed to be fixed.
Alternatively, one can say that an effective instanton measure
has a $\delta$-function peak in $\rho$.

(ii)  For the light quarks,
most  instanton amplitudes  are suppressed by powers of  the 
light quark masses. This suppression
is  due to the fermion zero modes of the Dirac operator, occurring in the
instanton-like backgrounds. We ignore these factors, as well as all
other pre-exponential overall factors coming from the instanton measure.
We  will only trace  dependences on large momenta
relevant to the problems under consideration ($Q$ in $e^+e^-$
annihilation,  the
heavy quark mass $m_Q$ in the case of the heavy quark decays, and so on).

(iii) We ignore all singularities of the correlation functions
at $x=0$. The corresponding contributions are associated
with the power terms in the condensate expansions.
The
 instantons-based models are presumably  not precise enough
to properly capture the condensates. We will isolate
and calculate only those contributions that come
from  the finite distance singularities
at
$x^2 = -\rho^2$.  In the momentum space
they produce  the exponential/oscillating
terms sought for.

\vspace{0.1cm}

It is seen that our evaluation of deviations from duality
is based on the most general aspects of the instanton formalism,
and, in essence,  does not depend on  details. 

The advantage of the model is its simplicity.
The only formula we will need  is
\beq
\int \, d^4 x \, \frac{1}{(x^2 +\rho^2)^\nu}\, e^{iqx} =
\frac{2\pi^2}{\Gamma (\nu )}
\left (\frac{Q\rho}{2}\right)^{\nu -2}
\frac{K_{2-\nu} (Q\rho  )}{\rho^{2\nu - 4}}
\label{basfo}
\eeq
in the Euclidean domain, which implies that in the Minkowski
domain, at large $q^2$,
\beq
{\rm Im} \, \int \, d^4 x \, \frac{1}{(x^2 +\rho^2)^\nu}\, e^{iqx} 
\propto 
s^{\frac{\nu}{2}-\frac{5}{4}}\, \sin (\sqrt{s}\rho -\delta)\,, \quad s=q^2\,.
\eeq
Here $K$ is the McDonald function, and $\delta$ is a constant phase which
is of no concern to us here.

To explain how it works it seems best to consider concrete examples.
Let us start from  
$e^+e^-$ annihilation. The polarization operator defined
in Eq. (\ref{Tproduct}) is given by the graph of Fig. 1.
In Sec. 2 we evaluated this graph for free quarks
and found that
$\rho (s) $ tends to a constant
 at asymptotically large energies.
Now, we evaluate the very same graph
using the quark Green functions in the background instanton field,
rather than the free quark Green functions. 
For massless quarks the Green functions $G_{\rm inst}$ are known exactly
\cite{Brown:1978eb}, but we will not need the full expression.
We will only need to know that $G_{\rm inst}$ is a sum of terms
of the following structure
\beq
G_{\rm inst} (x,y) = \frac{1}{[(x-y)^2]^2}\,
\frac{1}{[(x-z)^2+\rho^2]^{\ell_1}\, [(y-z)^2+\rho^2]^{\ell_2}}\, \tilde{G}\,,
\label{qgfprim}
\eeq
where $z$ is the instanton center, and
\beq
\ell_{1,2} = \frac{1}{2} \,\,\, {\rm or}\,\,\,\frac{3}{2}\,,\quad \ell_1+\ell_2 = 
2\,,
\label{qgf}
\eeq
and the numerator $\tilde G$ is a polynomial of $x,y$.
As was explained above, the singularity of $G_{\rm inst} (x,y)$ at
$x-y=0$ is irrelevant for the exponential/oscillating terms.
Of importance are the singularities in the complex plane coming from the
second factor in Eq. (\ref{qgfprim}). The integrals which one has  to take
can (and must) be evaluated at a saddle point;
 a simple analysis \cite{Chibisov:1997wf}
shows that at the saddle point the instanton center $z$ is exactly
in the middle between $x$ and $y$. Then the relevant singularities
of the quark correlation functions are  at $(x-y)^2=-4\rho^2$
(see  Eq. (\ref{qgfprim})). At the singularity  the first factor $1/(x-y)^4$
(which is singular at the origin)
can be replaced by $1/(16\rho ^4)$ and then safely omitted together with all
other prefactors.

We can proceed now to the calculation of the exponential/oscillating 
component of the polarization operator $\Pi_{\mu\nu}$.
The polarization operator is the product of two Green
functions (\ref{qgfprim}); therefore, at large Euclidean momenta
\begin{eqnarray}
\Pi_{\mu\nu} & \propto& \int \, d^4 x \,  e^{iq(x-y)} \, d^4 z
\frac{1}{[(x-z)^2+\rho^2]\, [(y-z)^2+\rho^2]^{3}}\nonumber\\[0.1cm]
& \propto& \int \, d^4 x e^{iqx}\, \frac{1}{x^2+\rho^2}
\times \int \, d^4 z e^{iqz}\, \frac{1}{(z^2+\rho^2)^{3}}\nonumber\\[0.1cm]
& \propto& K_1 (Q\rho )  K_{-1} (Q\rho ) \propto \frac{1}{Q}\,\exp (-2Q\rho )\,.
\label{ququ}
\end{eqnarray}
Note that once the integration over the instanton
center is carried out, the integral factorizes.

Equation (\ref{ququ}) implies that the exponential component of 
$\Pi (q^2)$, defined in Eq. (\ref{curco}),  is
\beq
\Delta \Pi \propto \frac{1}{Q^3}\,\exp (-2Q\rho )\, ,
\eeq
which implies, in turn, that at large $E=\sqrt{s}$ the oscillating component of
the spectral density is
\beq
\Delta\rho (s) \propto \frac{1}{E^3}\sin (2E\rho )\,. 
\label{homeone}
\eeq
Equation (\ref{homeone}) reproduces the high-energy asymptotics of
the exact result \cite{anca} for  the polarization operator
in the one-instanton approximation.

Does the $s^{-3/2}$ fall off of the oscillating (duality violating)
component make sense?  I will confront theoretical 
expectations with experiment in Sec. 11.
Here I just refer the reader to Fig. 11 (see the dashed curve),
postponing  a more detailed discussion till after both, the
instanton-based and the resonance-based models, are 
considered.

The next example to be analyzed is the total hadronic
$\tau$ width. This exercise is quite similar to previous exercises.
The corresponding transition operator is depicted in Fig. 4
(its imaginary part at $p^2 = M_\tau^2$ is proportional
to the width of the hadronic $\tau$ decay).  The neutrino 
Green function  clearly does not ``feel" the background
gluon field, so that the neutrino propagator is that of a 
free fermion. The same is valid for the $\tau$ lines. 
Therefore, we  immediately conclude that
\beq
\Delta\Gamma (\tau\to\nu +{\rm hadr.}) \propto \frac{1}{M_\tau}\sin
(2M_\tau\rho )\,,
\eeq
cf. Eq. (\ref{ququ}). The asymptotic (parton-model)
prediction in this case is $\Gamma (\tau\to\nu +{\rm hadr.}) \propto
M_\tau^5$, so that the oscillating component in $R_\tau$
scales as
\beq
\Delta R_\tau \propto \frac{1}{M_\tau^6}\sin (2M_\tau\rho )\,,
\label{homeon}
\eeq
where
\beq
R_{\tau} \equiv
\frac{\Gamma ( \tau ^- \to \nu _{\tau}+ {\rm hadrons})}
{\Gamma ( \tau ^- \to \nu _{\tau}e^- \bar \nu _e)}\,.
\label{rtau}
\eeq
\begin{figure}   
\epsfxsize=6cm
\centerline{\epsfbox{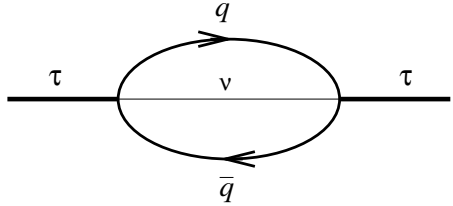}}
 \caption{The transition operator $\hat T$ determining the hadronic $\tau$
width, $\Gamma_{\rm hadr} (\tau ) = M_\tau^{-1}\,{\rm Im}\, \langle\tau |\hat
T |\tau\rangle$.}
\end{figure}

Note that the pre-exponential suppression factor in $\Delta R_\tau$
is significantly stronger than in  $\Delta R (e^+e^-)$, namely,
 $M_\tau^{-6}$ vs. $E^{-3}$. Of course,
to make quantitative statements it is not sufficient
to establish the scaling laws; one needs absolute normalizations.
Here, we are basically in uncharted waters.
At best, we have \cite{Chibisov:1997wf} some educated guesses,
which, if true, imply that
$\Delta R_{\tau} /R_{\tau} \lsim 5\%$. This estimate is rather close
to what one obtains for duality violations in $\tau$ in the 
resonance-based model, see Sec. 8.

In Sec. 1 I mentioned that the current interest to the
problem of the quark-hadron duality was driven, to a large extent,
by a significant progress in the experiments on the inclusive heavy
flavor decays. Theoretically, they can be treated along the same lines
as $e^+e^-$ annihilation or $\tau$ decays. The distinctions are technical.
Let us start from the total semi-leptonic width of the $b$ flavored hadrons.
At the quark level the process is described by the transition
$$
b\to q \ell^-\nu\,,
$$
where $q$ is either $u$ or $c$ quark, and $\ell$ stands for the electron, muon
or
$\tau$ lepton. We will first neglect the masses
of the final fermions; this is an excellent approximation for
$q =u$ and $\ell = e,\,\mu$. The impact of the final quark (lepton) masses
will be considered later.

The relevant transition operator is depicted in Fig. 5.
\begin{figure}   
\epsfxsize=6cm
\centerline{\epsfbox{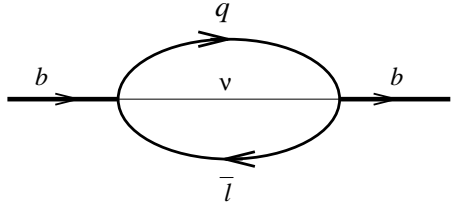}}
 \caption{The transition operator $\hat T$ determining the
total semileptonic width of $B$ mesons.}
\end{figure} 
The differences compared to the case of the $\tau$ decay are as follows:

(i) There is one (rather than two) light-quark line 
carrying a large external momentum;

(ii) Unlike the $\tau$ lines, the $b$ quark lines do experience
interactions with the background gluon field.

The analysis of the exponential/oscillating
component  in the heavy flavor decays combines
standard elements of the heavy quark expansion
(for a review see e.g.  \cite{Shifman:dn}), and the instanton calculus
(a pedagogical review can be found in \cite{ABC}). It is convenient 
to choose the rest frame of the heavy hadron at hand,
and single out the large ``mechanical" part in the $x$ dependence of the 
heavy quark field,
$$
Q(x) = e^{-im_Q t}\tilde Q (x)\,.
$$
Then the total width is proportional to the
imaginary part of the transition operator,
\beq
\Gamma = \frac{1}{M_{H_Q}}\langle H_Q |\hat T | H_Q\rangle\,,
\eeq
where
\beq
\hat T = i\int \, \bar{\tilde Q}(x) G (x,y) {\tilde Q}(y) D(x-y)
e^{im_Q(x_0-y_0)} d^4 (x-y) d^4 z\,,
\label{hometwo}
\eeq
$G (x,y)$ is the light quark Green function,
$ D(x-y)$ describes the propagation of colorless objects (the lepton 
pair in the case at
hand), while $z$ is the instanton center. The subscript 0 marks 
the time component.

Both, the  light quark Green function $G (x,y)$ and the heavy quark fields
$\tilde Q$ are to be considered in the background gluon field.
Taken separately, they are not gauge invariant;
the product  in Eq. (\ref{hometwo}) is.
In the leading order in the heavy quark expansion,
the heavy quarks propagate only in time; therefore,
\beq
\tilde Q(x_0,\, \vec x) = T e^{i\int_0^{x_0}A_0(\tau, \vec x)d\tau}
\tilde Q(0,\, \vec x)\equiv U(x) \tilde Q(0,\, \vec x)\, .
\eeq
It is convenient (although not necessary)
to impose the condition that at large distances from the
instanton center the quark propagation becomes free.
This condition implies that the singular gauge is used.
An explicit expression for $U(x)$ can be found in this gauge;
it is rather cumbersome, and I will  not quote it here,
referring the reader to the original publication
\cite{Chibisov:1997wf}. At the saddle point
(which again corresponds to the instanton situated
exactly   in the middle between the points, $x$ and $y$,
i.e. $z=(1/2)(x+y)$) the product $U^{-1} (x) ... U(y)$
in Eq. (\ref{hometwo}) reduces to unity, at least in the part 
which is singular
at $(x-y)^2 = -4\rho^2$. Note, that the matrices converting
the nonsingular-gauge Green function $G (x,y)$
to the singular gauge can be ignored too. This implies that
one may continue to use  Eq. (\ref{qgfprim}).

Thus,  the heavy quarks decouple from the
instanton background in the calculation of  the exponential/oscillating
terms. This is the consequence of
the fact that we exploit the heavy quark expansion and limit 
ourselves to the leading in $1/m_Q$ terms. In the subleading 
terms this decoupling does not necessarily take place.
Generally speaking,  the replacement of the $z$ integral 
by the value of the integrand at the saddle point
is not warranted in the next-to-leading orders.
This effect will not be further pursued, however.

Summarizing, the heavy quark expansion makes the heavy quarks sterile
with respect to the duality violating component. This is
certainly not counterintuitive. 
The duality violating component in the
heavy quark inclusive decays originates from the light-quark 
propagators. Its calculation reduces to  Eq. (\ref{qgfprim})
and the routine outlined after this equation (plus the basic formula (\ref{basfo})).
Starting from Eq. (\ref{hometwo})
we arrive at
\beq
\hat T \propto \frac{1}{m_Q^3}e^{-2m_Q\rho}
\label{phomethree}
\eeq
in the Euclidean domain,  which results in
\beq
\Delta \Gamma_{\rm sl} \propto \frac{1}{m_Q^3}\sin( 2m_Q\rho)\,.
\label{homethree}
\eeq
Since the total width scales as $\Gamma \propto {m_Q^5}$,
the oscillating component of the semileptonic branching ratio
is obviously suppressed by $ {m_Q^8}$,
\beq
\Delta {\rm Br} (H_Q\to \ell\nu +{\rm hadrons})\propto \frac{1}{m_Q^8}\sin
(
2m_Q\rho)\, .
\eeq

It is easy to generalize this formula to include an arbitrary number of the
light quarks in the final state. Each extra light quark adds two powers of $m_Q$
in the numerator. Thus,
in the total nonleptonic branching ratio, with three light quarks
in the final states,
\beq
\Delta {\rm Br} (H_Q\to  \mbox{
light hadrons})\propto \frac{1}{m_Q^4}\sin(
2m_Q\rho)\, .
\eeq
For the sake of
completeness I will also mention the radiative decays
of the type $b\to s+\gamma$. Assuming that these decays
are induced by local operators\footnote{In actuality
this is true only for a part of the amplitude.}
$$
\bar b \sigma_{\mu\nu} (1+\gamma_5) s F_{\mu\nu}\,,
$$
where $F_{\mu\nu}$ is the photon field strength tensor,
we immediately conclude that 
\beq
\Delta \Gamma (b\to s+\gamma ) \propto \frac{1}{m_Q^3}\sin( 2m_Q\rho)\,,
\eeq
precisely in the same way as in Eq. (\ref{homethree}).
The parton expression for $\Gamma (b\to s+\gamma )$ scales as
\beq
\Gamma_0 (b\to s+\gamma ) \propto m_Q^3\,,\quad m_Q\to\infty\,.
\eeq
As a result, our instanton-based model predicts that the duality
violating component in $b\to s+\gamma$ is suppressed as
\beq
\Delta \Gamma (b\to s+\gamma ) /\Gamma_0 (b\to s+\gamma ) 
\propto  \frac{1}{m_Q^6}\sin( 2m_Q\rho)\,.
\eeq

So far it was assumed that the quarks and leptons produced
were massless.
What changes would one decide to take into account
finite (nonvanishing) masses of the quarks and leptons?

The exact dependence on the masses of the final
quarks or leptons is rather sophisticated. Although it is calculable in principle,
the corresponding calculations are much harder to perform
than the simple estimates presented above. Moreover,
this is hardly necessary.  Given a crude
 nature of the model, which is intended for orientation
rather than for precision estimates, it
seems reasonable to treat 
$u,d$ and $s$ quarks 
as massless while $c$  as heavy, and the
$c$  lines as free propagation. Then the only impact of the
$c$ quark mass is kinematical, it results in the replacement of the total 
energy $m_Q$ by a relevant  energy
release in the light quarks in the process at hand. 

\begin{center}
$\star$ $\star$ $\star$
\end{center}

These are just a few  of  important applications which are
under discussion in the
current literature. The results are collected in Table 1.
Our instanton-based
model of the duality violation is user-friendly --
it is very easy to evaluate the exponential/oscillating
component in other inclusive processes not included in Table 1,
would such a necessity arise.

\begin{table}
\begin{center}
\begin{tabular}{|c|c|c|c|c|c|}
\hline
 ~  &~ & ~ & ~ & ~ & ~ \\[-0.1cm]
 $e^+e^-\to X$   &  $\tau\to\nu X$  & $H_Q\to \ell \nu X $  &$H_Q\to  X$ 
&$H_Q\to X\gamma$ & $H_Q\to H_{Q'} X$ \\[0.2cm]\hline
\vspace*{-0.2cm}
~ & ~  & ~  &~ & ~ & ~  \\
$\frac{1}{E^3}$ & $ \frac{1}{M_\tau^6}$  & $\frac{1}{M_{H_Q}^8}$ 
&$\frac{1}{M_{H_Q}^4}$ & $\frac{1}{M_{H_Q}^6}$ & $\frac{1}{\left(\Delta
M_{H_Q}\right)^6}$
\\[0.2cm] 
\hline
\end{tabular}
\caption{The index of the power fall-off of the oscillating (duality violating)
component in various inclusive processes, normalized to the corresponding
asymptotic (parton model) formulae, in the instanton-based model.
The capital $X$ denotes the light-quark hadronic states.}
\end{center}
\end{table}

\section{Global vs. Local Duality}

Usually by local duality people mean
 point-by-point comparison of $\rho (s)_{\rm theor}$ and
$\rho (s)_{\rm exp}$, while global duality compares
the spectral densities  $\overline{\rho (s)}$ averaged
over some {\em ad hoc} interval of $s$, with an 
{\em ad hoc} weight function $w(s)$,
$$
\int_{s_1}^{s_2} ds\,  w(s)\,  \rho (s)_{\rm theor} \approx
\int_{s_1}^{s_2} ds \, w(s)\, \rho (s)_{\rm exp}\,.
$$

Here I would like to comment on a very common misconception
which travels from one paper to another.
Many authors believe that  global duality defined in this way has a more solid
status than local duality. Some authors go so far as to say that
while global duality is certainly valid at high energies, this is
 not necessarily the case for local duality. This became a routine statement in the
literature. Well, routine does not mean correct.

In fact,
both procedures have exactly the same theoretical status.
The point-by-point comparison, as well as the comparison
of $\overline{\rho (s)}$'s (with an {\em ad hoc} weight function), must be
considered as distinct versions of local duality. 
The distinction between the ``local" quantities, such as   $R(e^+e^-)$ at a certain
value of
$s$ and the integrals of the type involved, say, in $R_\tau$  is 
quantitative rather
than qualitative. Comparison of Eqs. (\ref{homeone}) and (\ref{homeon})
makes this assertion absolutely transparent:
in both quantities there is a duality violating component,
the only distinction is a concrete index of the power fall-off
(3 vs. 6). 

The genuine global duality applies only to special integrals which can be 
{\em directly expressed through the
Euclidean quantities}. For instance, if the
integration interval extends from zero to infinity, and the weight function
is exponential, the integral
 $$
\int_{0}^{\infty} ds \exp\{-s/M^2\} \rho (s)\,,
$$
reduces \cite{SVZ} to the Borel transform of the
polarization operator  $\Pi (Q^2)$ in the
 Euclidean domain (i.e. at positive $Q^2$). For such quantities,
duality can{\em not} be violated, by definition.

There is one more aspect of  averaging (smearing)
which is not fully understood in the literature and requires comment.
Let us  consider again
$R_\tau$. It
can be expressed in terms of spectral densities
$\rho_V$ and $\rho_A$ in the vector and axial-vector
channels, respectively,
\begin{eqnarray}
R_{\tau}  \equiv
= \int _0^{M_{\tau}^2} \frac{{ d}s}{M_{\tau}^2}
\left( 1-\frac{s}{M_{\tau}^2}\right) ^2
\left( 1+2\frac{s}{M_{\tau}^2}\right)
\left[ \rho_V(s) + \rho_A(s)
\right] 
 \,.
\label{72}
\end{eqnarray}
(More exactly, the integration over $s$
runs from $M_\pi^2$, but in the chiral limit
we stick to, the pion mass vanishes.) The spectral densities
$\rho_V$ and $\rho_A$ are normalized in such
a way that their asymptotic
(free-quark) values are
$$
\rho_V(s)\, , \,\,\, \rho_A(s)\to N_c\quad\mbox{at}\quad s\to\infty\,.
$$
Correspondingly, the asymptotic value of $R_\tau$ is
$
R_\tau^0 = N_c
$.

In the instanton-based model, the duality 
violating component
in $\rho_{V,A}$ scales as $E^{-3}$. It is tempting to say then, that
the duality violating component in  $R_\tau$ can be obtained by
integrating the duality violating components of $\rho_{V,A}$
with the weight function specified in 
Eq. (\ref{72}). 

This would lead us nowhere, however. First of all,
the integral in Eq. (\ref{72}) runs all the way down to zero,
while Eq. (\ref{homeone}) is valid at asymptotically large $E$.
Even if the lower limit of 
integration were chosen to be high enough,
this would not help to find $\Delta R_{\tau}$ from Eq. (\ref{72}).
Indeed, $\Delta \rho_{V,A}$ is an oscillating function
of $s$, any smearing inevitably entails
cancelations, and the result would depend
on subtle details of $\rho (s) $, which are certainly beyond theoretical
control. The cancelations are seen, in particular, 
from the fact
that asymptotically  $\Delta R_\tau (M_\tau )$
falls off faster than $\Delta \rho_{V,A}(s)$, namely,
$M_\tau^{-6} $ versus $E^{-3}$.
There is no way one could predict the change of the index from
6 to 3 based solely on the integral (\ref{72}).

At the same time,
our instanton-based model does allow one to
predict the index in $R_\tau$. To this end one must
analyze the appropriate transition amplitude in the
Euclidean domain performing the analytic continuation
to the Minkowski domain at the very end.
The model captures enough intricate features of QCD to ``know"
the result of the smearing as a whole. Note that  
$\Delta R_\tau (M_\tau)$ depends on the highest scale in the problem
at hand, $M_\tau$, rather than on any intermediate or low 
scales one encounters
in process of integration in  Eq.  (\ref{72}). This is a general feature which
will be valid in any process and will persist
in any model based on evaluating
 singularities off the origin.

The case of 
$R_\tau$  is quite typical.
Similar questions arise in other problems,
where a similar strategy should be applied.
A problem of this
type which deserves a special mention in view of its
practical significance,
is that of the inclusive semileptonic decays
of $H_Q$. The decay rate can be obtained as an integral
over appropriate kinematic variables over the hadronic
structure functions,
\begin{eqnarray}
\Gamma(H_Q\to \ell\nu +X) &&=
|V_{qQ}|^2 \frac{G_F^2}{64\pi^2}\int dE_\ell \, dq^2\, dq_0
\nonumber\\[0.1cm]
&&\left\{2q^2 w_1 +\left[4E_\ell (q_0 - E_\ell )-q^2
\right]w_2 + 2q^2 (2E_\ell -q_0 )w_3
\right\}
\label{sldp}
\end{eqnarray}
where $q$ is the lepton pair momentum, $w_{1,2,3}$
are the hadronic structure functions which depend on $q_0$ and $q^2$
(for their
definition see e.g.  Ref. \cite{Blok:1994va}). 
Equation (\ref{sldp}) refers to the rest frame of $H_Q$.
Other notations are
self-explanatory. The exponential/oscillating contribution to the decay rate is
presented in Eqs. (\ref{phomethree}) and (\ref{homethree}), where it is
assumed that $m_q=0$. If one decided to get an estimate from Eq. (\ref{sldp}),
by integrating the duality violating contributions in the
structure functions, one would observe parametrically larger violations,
 exploding at the boundaries of the kinematically allowed domain.
These large violations cancel in the total rate
$\Delta \Gamma(H_Q\to \ell\nu +X) $, because of oscillations.

\section{A  Resonance-Based Model}

Now we  will  acquaint ourselves with another approach 
based on  the resonance saturation
of the colorless $n$-point functions. That this is a  
distinct dynamical source of  duality violations 
follows from the discussion in Secs. 3 and 5.
Shortly we will confirm this by observing that the functional
form of oscillations comes out different
compared to that in the instanton-based model. 
Theoretical analysis  is most transparent in the limit $N_c =\infty$.
 Later I will allow $N_c$ to become finite albeit large
(Sec. 10). We will first try to abstract general features, and then illustrate
them in the simplest possible dynamical setting which 
still exhibits confinement -- two-dimensional 't Hooft model.\cite{thooft} 

Let us first summarize what we expect to get for 
 the polarization operator defined in Eq. (\ref{Tproduct}) in the limit of large
$N_c$. In multicolor QCD,
$N_c\to\infty$, the resonance widths vanish. The spectrum of excitations in
the given channel 
is expected be (asymptotically) equidistant. A string-like picture
of the color confinement naturally leads to (approximately) linear Regge
trajectories. 
For each  primary trajectory there are infinitely many daughter
ones. The daughter trajectories are parallel
to the primary trajectory and are shifted by integers
(for a review see e.g. \cite{revven}). As a result, the excitation spectrum
in the given channel takes the form
\beq
M_n^2 = M_0^2 + \sigma^2 n\,,\quad \sigma^2 \equiv 2/\alpha '\,,
\label{rt}
\eeq
where $\alpha '$ is the slope of the trajectories. 
Note that the neighboring resonance states in the given channel
are
separated by the interval $ 2/\alpha '$ rather than
$ 1/\alpha '$. This is due to the alternating signatures of the
daughter trajectories. 

All these properties can be extracted from the  Veneziano amplitude found in
1968 (see Sec. 7.4 in Collins, \cite{revven})
which gave rise to the modern  string theory.

Equation (\ref{rt})  implies, in turn,  that  $\Pi (q^2 )$  can be presented
as  an infinite sum
\beq
\Pi (q^2 ) =-\frac{N_c \sigma^2}{12\pi^2}\,  \sum_{n=0}^\infty \frac{1}{q^2 -
M_n^2}= -\frac{N_c \sigma^2}{12\pi^2}\,\sum_{n=0}^\infty\, 
\frac{1}{q^2 - \sigma^2n- M_0^2}\,.
\label{resrep}
\eeq

\vspace{.2cm}

In the problem at hand there are two large quantities:
the number of colors and the energy. If one fixes $s$
and lets $N_c$ grow, one eventually arrives at the comb of infinitely
narrow $\delta$ functions, as in Fig. 6.  On the other hand,
if $N_c$ is fixed (no matter how large it is), with increasing $s$ one
eventually finds oneself in the energy region where the resonance
widths cannot be neglected; in fact, the resonances start overlapping,
and $\rho (s) $ gets smoothed.
The approximation of the infinitely
narrow resonances badly fails here, and must be amended.
These two limits  ($N_c\to \infty$ with $s$ fixed, and
 $N_c$ fixed with $s\to \infty$)  are not interchangeable.
Later I will include the nonvanishing widths (Sec. 10). For the time being
I put $N_c=\infty$, so that all resonance widths vanish. 

The  equidistant  spectrum only holds if both, the primary and all
daughter trajectories, are exactly linear and parallel.  This is 
not fully realistic.
Even putting $N_c=\infty$ does not help.
The low-energy parts of the
Regge trajectories in QCD are not exactly linear, in particular,
due to the spontaneous breaking of the chiral symmetry and the emergence
of the massless pions.  One can explicitly check that the radial excitations are
not quite  equidistant
in the 't Hooft model: they become equidistant
only asymptotically, at $n\gg 1$ ($n$ is the excitation number).
The situation in real QCD must be similar.
 
This shortcoming of the model  affects the condensate expansion
at low orders but has no impact on 
 deviations from duality at high energies. Since the phenomenon
under discussion  is related to
the  factorial divergence of the high-order terms,
letting the low-lying excitations ``breathe"  does not change the factorial
behavior \cite{Zhitnitsky:1996qa}, which is in one-to-one correspondence with
the spectral formula at large $n$, where the spacings $M_{n+1}^2-M_n^2$ must
be constant. Therefore, it is okay to use the linear pattern (\ref{resrep}).
 For the very same reason the residues of all resonances
in Eq. (\ref{resrep}) are taken to be equal. 
Fluctuations of the residues would show up in the condensate expansion;
they do not affect the estimate of the duality violations.
(For an additional remark on the residues
see Sec. 11.)

The infinite sum in Eq. (\ref{resrep}) reduces to a well-known
Euler's $\psi$ function, the logarithmic derivative of 
$\Gamma$,
\beq
\Pi (Q^2 ) =-\frac{N_c }{12\pi^2}\,  \left[
\psi (z) +\mbox{Const}\right]\,,\qquad
z \equiv \frac{Q^2+M_0^2}{\sigma^2}\,,
\label{hoho}
\eeq
(see e.g. \cite{sprav}). 
The irrelevant (subtraction) constant on the right-hand side is infinite, strictly
speaking. The occurrence of the $\Gamma$ function reminds us of
Veneziano's amplitude.

At positive values of $Q^2$ an asymptotic representation exists for
the
$\psi $ function,
\beq
\psi (z) = \ln z -\frac{1}{2z} - \sum_{k=1}^\infty
\frac{B_{2k}}{2k}\, z^{-2k}\, ,
\label{asypsi}
\eeq
where 
$B_{2k}$ stand for the Bernoulli numbers,
\beq
B_{2k} = (-1)^{k-1}\frac{2(2k)!}{(2\pi)^{2k}}\zeta (2k)\, ;
\label{bernou}
\eeq
here $\zeta$ is the Riemann function. (In some textbooks
$(-1)^{k+1}B_{2k}$ is called the $k$-th Bernoulli number 
and is denoted by $B_k$.) Equation (\ref{asypsi})
defines the asymptotic expansion of the polarization operator,
\beq
\Pi (Q^2)\, \stackrel{\longrightarrow}{_{_{Q^2\to \infty}}}\,
-\frac{N_c }{12\pi^2}\, \ln Q^2 + \sum_{k=1}^\infty \frac{C_k}{(Q^2)^k}
\,,
\label{pex}
\eeq
where the coefficients $C_k$ can be expressed through $B_{2k}$
(see Eq. (\ref{asypsi})) in a relatively straightforward manner.
The leading logarithmic term exactly coincides with the free
quark loop of Fig. 1 presented in Eq. (\ref{lqpi}). This explains our
choice of the resonance residues. The next-to-leading term is $1/Q^2$,
followed by higher power corrections. Equations (\ref{asypsi}) and 
(\ref{bernou}) highlight the
the factorial divergence of
the condensate series which I have already mentioned several times.

One might want to eliminate the $1/Q^2$ term from 
$\Pi (Q^2)$ to make the $1/Q^2$ expansion realistic --
 it is 
known \cite{SVZ} that,  in QCD with massless
quarks, the terms of the first order in $1/Q^2$ do not appear in 
$\Pi (Q^2)$. The
 power series starts from $\langle G^2\rangle /Q^4$
where $\langle G^2\rangle$ is the gluon condensate.
One could eliminate  $1/Q^2$, say,  by fine-tuning the parameter $M_0^2$.
If 
$
M_0^2 = \sigma^2/2 = 1/\alpha '
$,
 the $1/Q^2$ term cancels.
Although this might seem desirable, in fact this is hardly worth the bother
because the
model with exactly linear trajectories
is too rigid to be realistic anyway. 
The $1/Q^4$ correction will come out way too large.
To accommodate the gluon and mixed condensates properly
one would need  a more flexible model, with more than one adjustable
parameter. Therefore, I will feel free to simplify the
model further by putting $M_0^2=0$ and discarding the term with 
$n=0$ in Eq.~(\ref{resrep}). 

The spectral density corresponding to the infinite sum of the 
equidistant resonances is
shown in Fig. 6,
\begin{equation}
\rho_{\,V} (s) = \rho_A (s) = N_c \cdot \sum_{n=1}^\infty
\delta \left(\frac{s}{\sigma^2}
- n\right)\,;\qquad \sigma^2=\frac{2}{\alpha^{\prime}}\,.
\label{SPECTDENS}
\end{equation}

Let us truncate the power expansion (\ref{pex})
at some finite order, and examine the theoretical prediction for
$\rho (s) $ obtained from the Euclidean side.
It does not matter in which order we truncate.
Any  power term $(1/Q^2)^n$ is invisible in $\rho (s) $
at positive $s$:
analytically continuing to positive $q^2$ and taking the imaginary part
one ends up with the $\delta$ function and its derivatives.
The only imaginary part at positive $s$
comes from the analytic continuation
of $\ln Q^2$.  Thus, in the model at hand
$\rho (s)_{\rm theor} = 1$.
Comparing this with the comb of the $\delta$ functions
in Fig. 6 we conclude that the point-by-point duality is maximally violated:
between the resonances the spectral density is grossly
overestimated (the ``experimental" curve
runs  below the ``theoretical" expectation) while
at the peaks it is grossly underestimated
(the ``experimental" curve
runs  above the ``theoretical" expectation). 
What is most crucial, the deviations from duality do not
die off with energy. The power
index vanishes!
 
\begin{figure}   
\epsfxsize=9cm
\centerline{\epsfbox{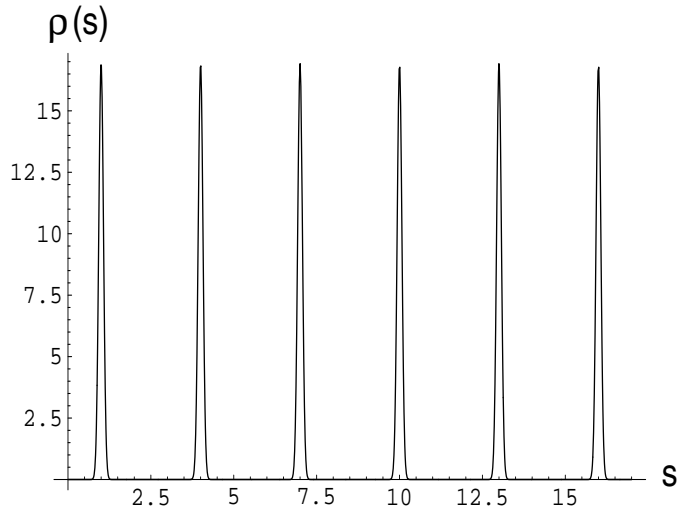}}
 \caption{The spectral density in the resonance model. For clarity I gave a
tiny width to the $\delta$ functions.}
\end{figure}

All this is pretty trivial. A somewhat less trivial
question  worth examining is the
impact of the {\em ad hoc} smearing.  For instance, let us average the spectral
density in Fig. 6 with the weight function appropriate to $R_\tau$, 
\begin{equation}
R_{\tau}  
=\frac{I_0(M_\tau^2)}{M_\tau^2} -3\,
\frac{I_2(M_\tau^2)}{M_\tau^6}+ 2\,\frac{I_3(M_\tau^2)}{M_\tau^8} \,,
\end{equation}
where the moments $I_n$ are defined as 
\begin{equation}
I_n(M)\;=\; \int_0^{M^2}\; {\rm d}s\, s^n\: \left[\rho_V(s)+\rho_A(s)\right]
\,.
\label{In}
\end{equation}
 To estimate the oscillating
contribution to $R_{\tau}$ which constitutes duality violation
that cannot be seen in a truncated OPE we treat  $M_{\tau}$
as a free (large) parameter. It will be seen momentarily
that
for $N_c = \infty$ and $M_{\tau}$ large, yet finite,
the duality violation in $R_{\tau}$ scales as
$1/M_{\tau}^6$. In other words, the vanishing power index in
$\rho$ translates\footnote{I remind that
the power index $\eta$ was defined in Sec. 3; in the case at hand
$\Delta R_\tau /R_\tau \sim M_\tau^{-\eta} \sin M_{\tau}^2/\sigma^2$. } in $\eta
= 6$ in
$R_\tau$. Certainly, this distinction is quantitative rather than qualitative, 
but, sure enough,
it is quite important  from the practical
side.

The sum over resonances in $R_\tau$ is easily calculated analytically:
for the spectral density of Eq.~(\ref{SPECTDENS}) it is
\begin{eqnarray}
R_{\tau} &\,=\,& R_{\tau}^{\mbox{\tiny OPE}} +\Delta R^{\rm osc}_\tau\,, 
\nonumber\\[0.2cm]
\frac{R_\tau^{\mbox{\tiny OPE}}}{N_c} &\,=\,& 1-\frac{\sigma^2}{M_\tau^2} +
\frac{1}{30}\left(\frac{\sigma^2}{M_\tau^2} \right)^4\,, \nonumber\\[0.2cm]
\frac{\Delta R^{\rm osc}_\tau}{N_c}&\;=\;& -\,
x(1-x)(1-2x)\, \left(\frac{\sigma^2}{M_\tau^2}\right)^3 \,+\,
\left[ x^2(1-x)^2-\frac{1}{30}\right]\,\left(\frac{\sigma^2}{M_\tau^2}\right)^4,
\label{n9}
\end{eqnarray}
where 
\vspace{0.2cm}
$$
x={\rm fractional~part~of}\left(\frac{M_\tau^2}{\sigma^2}\right), \;\;\;\: x\in 
[0,1)
\;.
$$

\begin{figure}
\epsfxsize=11cm
\centerline{\epsfbox{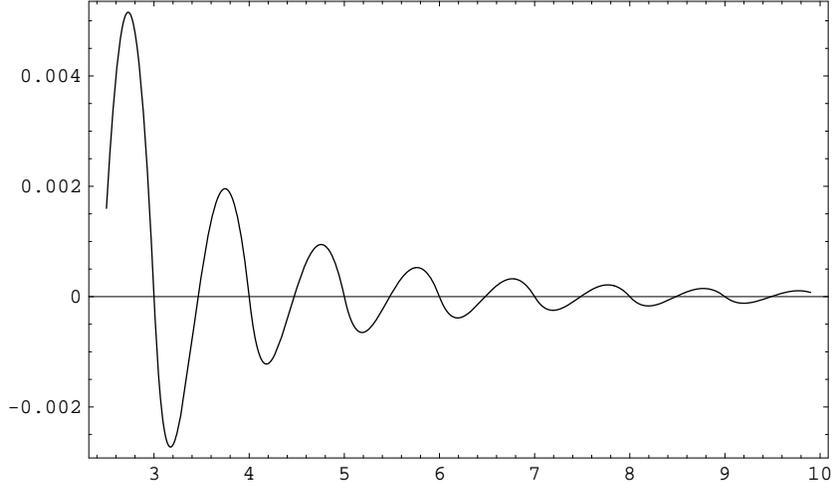}}
\caption{Oscillations in $R_\tau$. The plot of 
$\Delta R^{\rm osc}_\tau/R_\tau^0$  is presented as 
a function of  $M_\tau^2/\sigma^2$.}
\label{oscillation}
\end{figure}

\vspace{0.2cm}
I presented the result as a sum of two functions of $M_\tau^2$.  The first 
one, $R_{\tau}^{\mbox{\tiny OPE}}$, is a smooth function expandable in
$1/M_\tau^2$.   The second one,
$\Delta R^{\rm osc}_\tau$, oscillates with the period $\sigma^2$; its average 
vanishes, see the plot of $\Delta R^{\rm osc}_\tau/R_\tau^0$ in
Fig.~\ref{oscillation}. Although $\Delta R^{\rm osc}_\tau$ is not a pure sine
-- it contains higher harmonics -- the coefficients of the
higher harmonics are numerically suppressed. With the accuracy of 
 a few percent one can write
\begin{equation}
 \frac{\Delta R^{\rm osc}_\tau}{R_\tau}\;=
-\frac{1}{3\sqrt{12}}\, \left(\frac{\sigma^2}{ M_\tau^2}\right)^3
\sin{\left(2\pi\frac{M_\tau^2}{\sigma^2}\right)}
\, .
\label{n13}
\end{equation}

I pause here to make a comment.
The contribution of any particular resonance of mass $M_k$ to $R_\tau$,
according to Eq.~(\ref{72}), is given by a simple polynomial in
$1/M_\tau^2$ (times the step function $\theta(M_\tau^2-M_k^2)\,$).
Variations of parameters of a given resonance  
(or resonances)  change only the regular terms of
the $1/M_\tau^2$ expansion, but have  no impact on the 
oscillatory component.
From Eq.~(\ref{72}) it is clear that such variations change 
only coefficients of the
$1/M_\tau^2$, $1/M_\tau^6$ and
$1/M_\tau^8$ terms.   OPE for
$R_\tau$ must exactly reproduce these three expansion 
coefficients, and so it does.

In fact, it is not difficult
to demonstrate that $R_{\tau}^{\mbox{\tiny OPE}}$ coincides
with the OPE  prediction in the model at hand. The
power corrections can be
presented as follows:
\begin{equation}
R_\tau^{\mbox{\tiny OPE}}\;=\; N_c\,+\, \frac{\tilde I_0}{M_\tau^2}
\,-\, 3\frac{\tilde I_2}{M_\tau^6}
\,+\, 2\frac{\tilde I_3}{M_\tau^8}\;,
\label{n1}
\end{equation}
where the ``condensates'' $\tilde I_n$ are
\begin{equation}
\tilde I_n\;=\; 
\int_0^{\infty}\; {\rm d}s\, s^n\: \left[\rho_V(s)+\rho_A(s)-2N_c \right]
\;.
\label{tildIn}
\end{equation}
These integral representations for the ``condensates'' $\tilde I_n$ follow from
Eqs.~(\ref{72}),~ (\ref{In}) if one assumes that the spectral densities approach 
their 
asymptotic limits faster than any power of $1/s$. In the model at hand, with
 the comb-like spectral density,  the integral
representation (\ref{tildIn}) requires  regularization.
As a regularization one can introduce the weight factor $\exp(-\epsilon s)$, 
taking
the limit  $\epsilon \to 0$ at the end. With this regularization, 
$R_\tau^{\mbox{\tiny OPE}}$ from Eq.~(\ref{n9}) is reproduced.

The same result for the coefficients of the power terms in
$R_\tau^{\mbox{\tiny OPE}}$ could be obtained directly
 from the expansion of $\Pi (Q^2)$ in  Eq. (\ref{hoho}). We agreed to put
$M_0^2=0$ for simplicity. Then the expansion coefficients of $\Pi (Q^2)$
are those of the $\psi$ function in Eq. (\ref{asypsi}).
To find the power terms in $R_\tau$ one may analytically continue
the power terms in Eq. (\ref{asypsi}) to Minkowski, take the imaginary part
and convolute with the weight function presented in Eq. (\ref{72}).
It is easy to see that the only relevant terms  in Eq. (\ref{asypsi}) are 
$1/z$ and $1/z^4$ giving rise to $\delta (s)$ and $\delta ''' (s)$ in 
the imaginary
part. The terms of the higher order in $1/z$ drop out
because the weight function is a polynomial of the third order;
it contains no $s^4$  or higher terms. The term $1/z^2$ in Eq. (\ref{asypsi})
(it would generate  $\delta '(s)$)
drops out because the weight function does not contain  terms linear in $s$.
Then, the expansion for $R_\tau^{\mbox{\tiny OPE}}$ is  immediately 
recovered provided that one substitutes the appropriate value for
the appropriate Bernoulli number,  $B_2 = -1/30$.

A few words on the numerical aspect.
 Our consideration is admittedly illustrative. One
should not take too literally the numbers which ensue for many reasons: 
in particular,  $M^2\tau$  is not much larger than the spacing between the
resonances,  $N_c=3$ is probably not large enough to warrant the zero
width approximation, and so on. I would not put too
much confidence on particular numbers. 
I would settle on the statement that the above estimate of the
oscillation component is valid, say, up to a
factor of two or so. 
Taking our formula for $\Delta R^{\rm osc}_\tau$  at its
face value and  using the actual value of the 
$\tau$ mass ($M_\tau^2/\sigma^2 \sim 1.5$)
we obtain that
$\Delta R^{\rm osc}_\tau/R_\tau\sim
3\%$. It is rather rewarding to see that this estimate
is in the same ball park as that obtained in the 
instanton-based model. It seems safe to conclude that
duality violations in the total hadronic $\tau$ width
are expected at a level of a few percent. We do not 
know whether two
mechanisms add constructively or destructively.
Given this additional uncertainty, it will be no
 exaggeration
to assert that the overall  theoretical uncertainty 
$$\Delta R_\tau/R_\tau =
3\%\,.$$
The 3\% uncertainty in the hadronic $\tau$ width
translates into $\sim$20\% uncertainty in
$\alpha_s (M_\tau )$, which entails the uncertainty
of about 6\% in $\alpha_s (M_Z )$.

\begin{center}
$\star$ $\star$ $\star$
\end{center}
In  summary:

\vspace{0.1cm}

$\bullet$  The resonance structure of the hadronic spectrum
associated with the confining properties of QCD
leads to a distinct exponential/oscillating component 
invisible in the truncated OPE. 

$\bullet$ This component is related to  singularities
of the appropriate  $n$-point functions at infinite separations 
(in the coordinate space).

$\bullet$ The corresponding duality violations
are maximal at $N_c =\infty$, when the resonance widths vanish.
The power index which was introduced in Sec. 5
is calculable in many instances.
Generally speaking, the power indices
obtained in the instanton-based and resonance-based
models  do not coincide with each other. $R_\tau$
seems to be an exceptional case. We do not understand the reasons
explaining the coincidence of the  power indices in $R_\tau$;
probably, it is accidental.

$\bullet$ Inclusion of the nonvanishing
resonance widths will replace the power suppression in the 
pre-exponent by a
weak exponential suppression, see Sec. 10. The change of the 
regimes will probably have little numerical  impact in the $\tau$ decays
since $M_\tau^2$ is not large enough
for the exponential regime to develop in earnest.
The factor $(\sigma^2/M_\tau^2)^3$ in Eq. (\ref{n13})
will be replaced by
$$
{\exp}\left(
- \frac{2\pi B M^2_{\tau}}{N_c\sigma^2}
\right)\,.
$$
Both factors are rather close numerically.

\section{Numerical Illustrations in the 't Hooft Model}

The general pattern established in Sec. 8 is nicely illustrated
by numerical calculation in the 't Hooft model
which is ideally suited for exhibiting the resonance mechanism
of duality violations.
Indeed, in  the 't Hooft
model \cite{thooft} (two-dimensional quantum chromodynamics
considered in the limit $N_c\to\infty$, while $\alpha_s N_c =$ const)
the color confinement is automatic because  the
Coulomb potential in 
$1+1$ dimensions grows linearly with distance. 
For high excitation numbers the meson spectrum grows linearly,
$M_n^2 \propto n$. 
In two dimensions the gluon field is a nondynamical degree of freedom,
there are no transverse gluons, and the only remnant
of the gluon field is the Coulomb (instantaneous)
potential. There are no instantons, therefore,  the mechanism
of Sec. 6 does not overshadow the resonance mechanism.
Moreover, the quark loops are
suppressed at  $N_c\to\infty$ --  hence,
all quark mesons are infinitely narrow.
The spectrum is calculable, albeit numerically, and so are exclusive
decay amplitudes, which can be then summed up, one by one.
This replaces the experimental data, to be compared with 
the inclusive
OPE-based calculations. In a sense, this is even  better
than real data in actual QCD which are always 
incomplete and imprecise. In the 't Hooft model
one can make  {\em gedanken} experiments as
 complete and precise as one
wishes.

A thorough analysis of the
quark-hadron duality in the semileptonic heavy flavor
decays  in the 't Hooft model was
carried out by  Lebed and Uraltsev
 \cite{Lebed:2000gm}. Below I will present their result, but at first
let me remind relevant  aspects of the  't Hooft model.

The dynamical mass scale in the model is set up by the gauge coupling 
constant $g$ (which is dimensionful in $D=2$),
\begin{equation}
\beta^2 \equiv \frac{g^2}{2\pi} (N_c - 1/N_c) \, .
\label{ustalone}
\end{equation}
 Thus, $\beta$
plays the role of $\Lambda_{\rm QCD}$.  
Being finite at $N_c\to\infty$, it  provides a
natural unit of mass.  The quarks with masses less than
$\beta$ can be called light, while if $m_Q\gg \beta$
we are dealing with a heavy quark. To make contact with real QCD the
heavy quark will be
referred to as the $b$ quark, while the corresponding 
$ b\bar q$ bound state as the $B$ meson.

The semileptonic widths $b\to q e\bar\nu$
are induced by the  weak decay Lagrangian
\begin{equation}
{\cal L}_{\rm weak} \,=\, -\frac{G}{\sqrt{2}}\,
(\bar q \gamma_\mu b)
\,(\bar{e}\gamma^\mu \nu )\,,
\label{120}
\end{equation}
where $G$ is a ``Fermi constant," and it is assumed that
\begin{equation}
m_e = m_\nu =0\,.
\label{lmv}
\end{equation}
Note that in $D=2$ the axial current is not independent, so
that ${\cal L}_{\rm weak}$ is chosen to be 
pure  vector times vector. Moreover,  the
invariant mass $k^2$ of the lepton pair is always zero \cite{Bigi:1998kc},
provided the lepton masses vanish, Eq. (\ref{lmv}). 
This fact is specific for 2D models; it
implies that in two dimensions the semileptonic decays
$b\to q e\bar\nu$  are
equivalent to decays into a single massless pseudoscalar particle
$\phi$ weakly coupled to quarks,
\begin{equation}
{\tilde{\cal L}}_{\rm weak} = -\frac{G}{\sqrt{2\pi}}\, 
\bar q \gamma_\mu b \, \epsilon^{\mu\nu} \,\partial_\nu \phi \,.
\label{122}
\end{equation}

For simplicity it will be assumed that
the quark $q$ in the transition $b\to q e\bar\nu$ is
distinct from the spectator light antiquark in the $B$ meson, so
that annihilation diagrams are absent.
The leading (parton-model) result for the semileptonic width
$\Gamma (b\to q e\bar\nu)$
scales as $m_b^1$ (to be compared with $m_b^5$
in real QCD). A few first terms in the operator product expansion
for $\Gamma $  are
rather trivial;
it is not difficult to obtain
\begin{equation}
\Gamma_{\rm OPE}= \frac{G^2}{4\pi} \cdot \frac{m_b^2 -
m_q^2}{m_b} \cdot \frac{m_b}{M_B} \int_0^1 \frac{{\rm d}x}{x}
\,\varphi_B^2(x)\, .
\label{141}
\end{equation}
up to terms  terms $O(m_b^{-4})$. Here $x$ is the $b$
quark momentum fraction of $B$ in light-cone coordinates,
while $\varphi_B(x)$ is the light-cone wave function.
The expectation value 
$(m_b/M_B) \left\langle 1/x \right\rangle$ 
in Eq. (\ref{141}) can, in turn,  be computed in the form 
of a $1/m_b$ expansion.

Equation (\ref{141}) presents
a smooth function of $m_b$, a theoretical part of the duality relation.
It carries no direct information
on the opening of new thresholds
each time $M_B$ crosses the successive
values of $M_n$ where $M_n$ is the mass of the $n$-th excited state 
in the $\bar q q$ channel (I will denote the meson itself
by the same symbol, $M_n$).

The ``experiential" part is obtained as a sum
over individual exclusive  decay channels of the type
$B\to M_n e\bar\nu$. Sure enough,  the ``experiential" part
includes the effects of the kinematic thresholds
in the most straightforward manner.
The spectrum and the light-cone  wave functions 
of the excited light mesons are found, numerically, from the
't Hooft equation \cite{thooft}. The concrete calculations
I would like to quote are done \cite{Lebed:2000gm}
at $m_q =0.56\beta$. The $b$ quark mass in these calculations
varies from $m_b = \beta$ (unrealistically light,
no phase space for excitations at all)
up to $m_b =12\beta$, when the highest kinematically allowed
excited  $\bar q q$ meson has the excitation number 18.
($12\beta$ corresponds, roughly speaking, to
$m_b = 4.5$ GeV in actual QCD.) The result \cite{Lebed:2000gm}
for
$$
\frac{\left(\Gamma_{B}\right)_{\rm  ``exp"}}{\Gamma_{\rm OPE}} -1
$$
is shown in Fig. 9, as   function of the $b$ quark mass. 

\begin{figure}
\epsfxsize=11cm
\centerline{\epsfbox{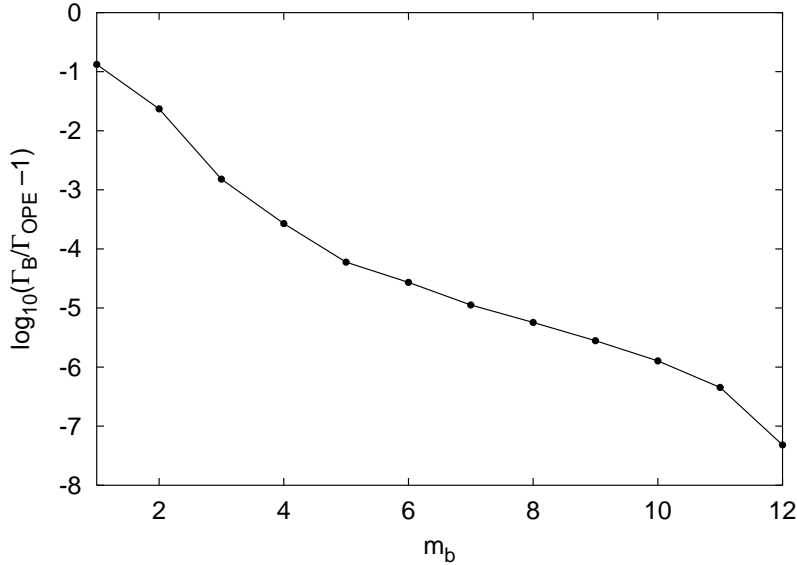}}
\caption{Deviations from duality
in the semileptonic $B$ decays in the 't Hooft model,
from Ref. \cite{Lebed:2000gm}. The $b$ quark mass
is measured in the units of $\beta$, the light quark masses are set at
$0.56\beta$.}
\label{LU}
\end{figure}

Seeds of oscillations inherent to duality violation  are clearly seen.  
(The deviations do
not average to zero but rather oscillate around a rapidly dissipating
contribution
which can be  attributed to discarded higher-order OPE terms.)
Numerically, the duality violating component turns out to be
extremely small -- almost certainly, an artifact of the 't Hooft model.
What I would like to emphasize is not the numerical suppression
 {\em per se}, but a general character of the phenomenon.

\section{The Impact of the Resonance Widths}

As was already mentioned, no matter how large $N_c$ is, 
at sufficiently high energies
one cannot neglect the resonance widths. One 
should also remember that in the real world the number of colors is
 three --
not very large by any count. 
One can expect 
that the nonvanishing resonance widths, once the   resonances start to overlap,
provide an additional  suppression of the duality violating component.

Below I will show that, inside the domain of overlapping,
the power regime (\ref{SSS})
is replaced by an exponential one, see Eq. (\ref{SSSS}),
even if the power index vanishes (as is the case for the
spectral density (\ref{rhoim}) presented in Fig. 6). 
The basic idea is that the  nonvanishing resonance widths
shift the poles away from the physical cut, to unphysical sheets,
which automatically results in a smoother  imaginary part on
the physical cut,  much closer to that obtained from OPE.

I return to the equidistant resonances  
considered in Sec. 8
in the leading order in $1/N_c$, with the intention to include
the next-to-leading $1/N_c$ effects. For any given excitation $n$,
the widths scale as $\Gamma_n /M_n\sim
N_c^{-1}$. Below we will analyze an infinite
 sequence of resonances
for which we will
 need to know the dependence of this
ratio on the excitation number $n$. For the time being, however,
 let us focus on
an ``isolated" resonance. 

In the leading order its contribution to the
polarization operator $\Pi (q^2)$ is
\beq
\frac{g_n^2}{q^2 - M_n^2}\, ;
\label{Wedone}
\eeq
here $g_n^2$ is the residue.
If $\Gamma_n\neq 0$, in the vicinity of the pole one can use the Breit-Wigner
formula which replaces Eq. (\ref{Wedone}) by
\beq
\frac{g_n^2}{q^2 - M_n^2 +i\Gamma_n M_n}\, .
\label{Wedtwo}
\eeq
Strictly speaking, $g_n^2$ and $M_n^2$ in 
Eqs. (\ref{Wedone}) and  (\ref{Wedtwo})
are different: in the former the residue and the mass must be
taken in the leading order in $1/N_c$ while in the latter
they include $O(1/N_c)$ corrections. This effect is less important
than that due to  $\Gamma_n$'s.  In many instances
(but not always, see below)
the $1/N_c$ shifts  of $g_n^2$ and $M_n^2$ can be neglected. 

Now comes a crucial assertion.
At large $n$ the resonance width
must scale as $\Gamma_n \sim M_n/N_c$, i.e.
the $n$ dependence of $\Gamma_n$ is the same as that of  $M_n$,
the square root of $n$.
This behavior was predicted long ago \cite{Nuss} 
on the basis of a simple qualitative picture;
much later it was confirmed \cite{Blok:1998hs} by numerical studies in
the 't Hooft model. This formula
 can be explained as follows. For highly excited states a
quasiclassical treatment applies. When a meson is 
created by a local source, it can
 be considered, quasiclassically, as a pair of (almost free) 
ultrarelativistic quarks; each of them with energy $M_n/2$. 
These quarks are  produced
at the origin, and then fly back-to-back, creating behind them a flux 
tube of the chromoelectric field. The length of the tube $L\sim 
M_n/\Lambda^2$
where $\Lambda^2$ represents the string tension. 
The meson decay width is 
determined, to order $1/N_c$, 
 by the probability of producing an extra quark-antiquark pair. Since 
the pair creation 
can happen anywhere inside the flux tube, one  expects 
that\footnote{Let me note in passing that the $1/N_c^2$ corrections due to 
creation of two quark pairs are of order $L^2/N_c^2$ within this 
picture. Since $L\sim M_n \sim \sqrt{n}$, the expansion parameter is
$\sqrt{n}/N_c$. } $\Gamma_n\sim L\Lambda^2/N_c$. Taking into 
account 
that $L\sim 
M_n/\Lambda^2$ one arrives at
\beq
\Gamma_n=\frac{B}{N_c}M_n\,,
\label{thurone}
\eeq
where $B$ is a dimensionless coefficient of order one. 
Naively extrapolating Eq. (\ref{thurone}) to $n=1$ and $N_c=3$
(well beyond the limits of its applicability)
and normalizing by  the $\rho$ meson for which $\Gamma /M \sim 0.2$,
one can make an educated guess,
\beq
B\sim 0.5\,.
\label{thurtwo}
\eeq

In the 't Hooft model, in which both
the meson masses and widths are calculable,
one can explicitly test the formula $\Gamma_n \sim \sqrt{n}$. 
It is curious that in
the 't Hooft mode one arrives
 at a close numerical
estimate for $B$.  Figure 9 shows $\Gamma_n$ versus $n$.
The width of the $n$-th excitation is computed numerically,
by summing up all open two-meson decay channels,
$$
\Gamma_n =\sum_{b,c} \Gamma (n\to b+c)\,,\quad M_b+M_c\leq M_n\,.
$$
The necessary three-meson constants are found
\cite{Blok:1998hs} as appropriate overlap integrals of the light-cone wave
functions, the calculation of $\Gamma_n$ extends up to 
the excitation number 500. It is seen that the solid curve yielding
 the $\sqrt{n}$
dependence fits well the numerical data.\footnote{Seemingly random 
fluctuations of $\Gamma_n/ \sqrt{n}$ around a constant value
are also clearly seen. A theory of these fluctuations
might be a good exercise. Developing such a theory might be fun.}
  Since $M_n \sim \sqrt{n}$, the
scaling behavior (\ref{thurone}) is confirmed. The coefficient $B$ defined 
in Eq. (\ref{thurone}) is obtained from the coefficient $A$ of Ref.
\cite{Blok:1998hs} as follows: $B= A/\pi^3$. Since $A\sim 15$
(see the Erratum), we recover Eq. (\ref{thurtwo}). 
 
\begin{figure} 
\epsfxsize=11cm
\centerline{\epsfbox{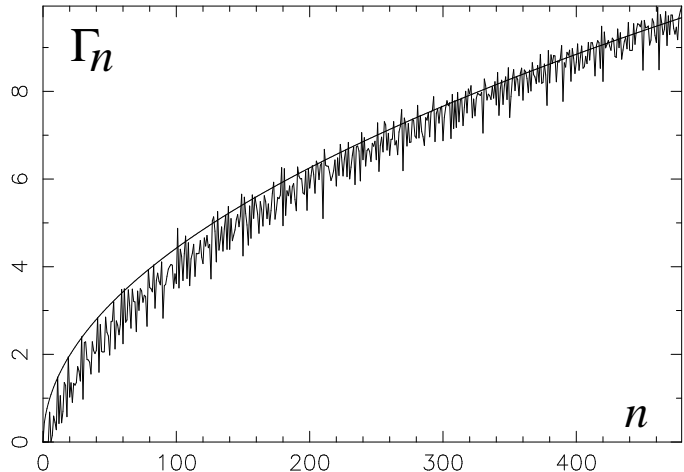}}
\caption{The width of the $n$-th excited meson in the 't Hooft model,
in the units of $32 \pi^{-2}N_c^{-1}\,\beta$, from 
Ref. \cite{Blok:1998hs}. Note that $M_n = \pi \beta \sqrt{n}$
at $n\gg 1$. The parameter $\beta$ is defined in Eq. (\ref{ustalone}).}
\label{BSZ}
\end{figure}

The Breit-Wigner resonance formula now takes the form (in the Euclidean
notation)
\beq
-{g_n^2}\left[ Q^2 + M_n^2 - i\frac{B M_n^2}{N_c}\right]^{-1}\, .
\label{thurthree}
\eeq
One must exercise certain care in using the Breit-Wigner expression
for the resonance contribution
in the polarization operator far away from the pole.
It must be adjusted in such a
way that the analytic properties of $\Pi (q^2)$ are not spoiled.
Namely, $\Pi (q^2)$ must remain analytic everywhere in the complex
$q^2$ plane, with a cut along the positive real semi-axis.
No singularities are allowed on the physical sheet, all poles
must be shifted to unphysical sheets. This is relatively 
easy to achieve. To this end let us replace Eq. (\ref{thurthree})
by 
\beq
-{ \tilde g_n^2}\left[ Q^2 \left(1-\frac{B}{\pi N_c}\ln \frac{Q^2}{\sigma^2}\right) +
\tilde M_n^2
\right]^{-1}\, .
\label{thurfour}
\eeq
On the upper side of the physical cut near the pole (i.e. at $Q^2$ in 
the vicinity of $-M_n^2 -i \epsilon$) the imaginary parts of
both expressions (\ref{thurthree}) and (\ref{thurfour})
coincide, 
provided that 
$\tilde g_n^2$ and $\tilde M_n^2$ are appropriately
adjusted ($\tilde g_n^2$ and $\tilde M_n^2$ differ from 
$g_n^2$ and $M_n^2$ by a $1/N_c$ correction;
in what follows we will omit the tilde, 
making the adjustment at the very end). Thus, both expressions are
equally legitimate for the Breit-Wigner description of the 
resonances. Moreover, Eq. (\ref{thurfour}), in turn, can be replaced by
\beq
-{  g_n^2}\left[ Q^2 \left( \frac{Q^2}{\sigma^2}\right)^{-\frac{B}{\pi N_c}} +
 M_n^2
\right]^{-1}\, .
\label{thurfive}
\eeq
The distinction between Eqs. (\ref{thurfour}) and (\ref{thurfive})
is of the order $O(1/N_c^2)$. We do not track such terms anyway.
This latter formula has the required analytic properties --
it is nonsingular everywhere in the complex   $Q^2$ plane, 
 except the cut at negative real $Q^2$. On the physical sheet of $Q^2$,
the variable
\beq
z\equiv   \left( \frac{Q^2}{\sigma^2}\right)^{1-\frac{B}{\pi N_c}}
\label{thsix}
\eeq
never becomes real and negative, so that the 
 pole singularities in Eq.  (\ref{thurfive})
are indeed shifted to  unphysical sheets. 

Summing over the infinite chain of the equidistant resonances, as in Eq. 
(\ref{resrep}),  we arrive at
\begin{eqnarray}
\Pi (Q^2) &=&\frac{N_c\sigma^2}{12\pi^2}\,\,
\frac{1}{1-{B}/(\pi N_c)}
\, \sum_{n=1}^\infty \left[ Q^2  \left(
\frac{Q^2}{\sigma^2}\right)^{-\frac{B}{\pi N_c}} +
 M_n^2
\right]^{-1}
\nonumber\\[0.3cm]
&=& -\frac{N_c }{12\pi^2}\,  
\frac{1}{1-{B}/(\pi N_c)}\, 
\left[
\psi (z) +\frac{1}{z}\right]\,,
\label{frione}
\end{eqnarray}
where $z$ is defined in Eq. (\ref{thsix}), and  $1-{B}/(\pi
N_c)$ in the denominator 
reflects the adjustment of the residues discussed above (this factor can be
established by demanding the correct asymptotic behavior
at $Q^2\to\infty$, cf. Eq. (\ref{lqpi})).

How does this compare with the zero-width approximation of Sec. 8?
Formally both results for the polarization operator
look similar; the only difference\footnote{The $1/z$
term which was absent in Sec. 8 is due to the fact that I put
$M_0^2 =0$ and start the summation from $n=1$ rather than from $n=0$.
These simplifications resulting in the
occurrence of  the $1/z$ term,
are irrelevant in the studies of the
duality violating component at high energies.}
 is in the definition of the variable $z$. In the deep Euclidean domain the 
power expansion of $\Pi (Q^2 )$ ensues from the
asymptotic representation (\ref{asypsi}). The $k$-th term
of the expansion, 
which in the leading order in $1/N_c$  used to be
$$
\frac{B_{2k}}{2k}\left(\frac{\sigma^2}{Q^2}\right)^{2k}
\,\,\, \qquad 
\mbox{(zero-width approx.)},
$$
now becomes
\beq
\frac{B_{2k}}{2k}\left(\frac{\sigma^2}{Q^2}\right)^{2k}\left(1
+\frac{2k B}{\pi N_c}\ln\frac{Q^2}{\sigma^2} +
\frac{B}{\pi N_c}+ O\left(\frac{1}{N_c^2}\right)
\right)\,. 
\label{fritwo}
\eeq
The impact of the next-to-leading $1/N_c$ terms on the power 
expansion is quite insignificant. In particular, the
$(B/N_c)\ln Q^2$ correction mimics the logarithmic 
anomalous dimension typical of OPE in QCD. 

At the same time, the $1/N_c$ effects radically affect
Im$\Pi$ on the physical cut (the spectral density).
Instead of the comb of the $\delta$ functions of Fig. 6, we now get
at high energies  a smooth function, with  mild oscillations. The plot of
the spectral density
$\rho (s) = (12\pi/N_c)\, {\rm Im}\Pi$
(with $\Pi$  determined by Eqs.
(\ref{frione}) and (\ref{thsix})) is presented in Fig. 10.

\begin{figure}   
\epsfxsize=10cm
\centerline{\epsfbox{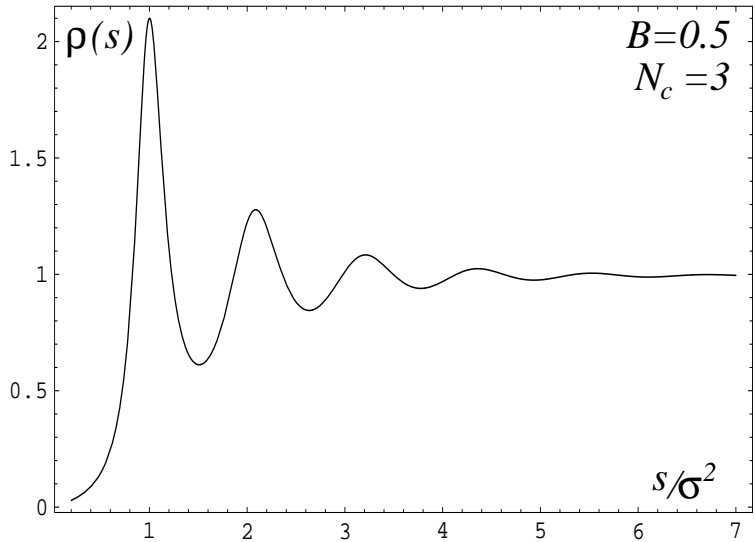}}
 \caption{The spectral density for
the  equidistant resonances with finite widths specified in Eq. (\ref{thurone}).}
\end{figure}

At asymptotically large $s$ the spectral density tends to unity,
plus power corrections corresponding to OPE
plus an oscillating component which cannot be obtained in
the power expansion. One can readily isolate it
by exploiting the well-known reflection property of Euler's
function,
\beq
\psi (z) +\frac{1}{z} =\psi (-z) -\pi {\rm cot} \pi z\,. 
\label{frithree}
\eeq
On the physical cut (in units of $\sigma^2$)
\beq
z= - s^{1-\frac{B}{\pi N_c}}\left( 1 + i\frac{B}{N_c}\right) +
O\left(\frac{1}{N_c^2}\right)\, ,
\label{frifour}
\eeq
and the imaginary part of the left-hand side
of Eq. (\ref{frithree}) is essentially given by that
of $-\pi {\rm cot} \pi z$. The imaginary part of
$\psi (-z)$ gives rise to non-oscillating power corrections $s^{-2k}$
in  the spectral density corresponding to OPE. Alternatively, 
these OPE corrections can  be obtained
 by a direct analytic continuation
to Minkowski
of the power terms (\ref{fritwo}).

As a result, the finite-width resonance model leads
us to
\beq
\rho (s) \to 1 +\mbox{power corr.} + 2\exp\left( {-\frac{2\pi s B}{\sigma^2 N_c}}
\right)\, 
 \cos \left(\frac{2\pi s }{\sigma^2 }
\right)\,,
\label{monone}
\eeq
where it is assumed that 
\beq
\frac{2\pi s B}{\sigma^2 N_c}\gg 1\qquad \mbox{but}\qquad \frac{\ln s}{N_c} \ll
1\,.
\label{montwo}
\eeq

If $s$ is fixed while $N_c$ is set to $\infty$ we recover
unsuppressed oscillations, as in Sec. 6.
In the opposite limit $N_c$ fixed and $s$ large we observe an exponential
suppression, with a weak exponent proportional to  $1/N_c$. 

A key question one can ask in connection with
the above analysis is as follows.
For the given value of $N_c$, what is the boundary energy marking the onset of
the exponential suppression? The best I can say at the moment is
that this boundary energy scales as $s_0 \propto N_c$. A reliable 
determination of
the proportionality coefficient  is a task for the future.  A naive estimate
following from Eq. (\ref{montwo}),
$$
s_0\sim \frac{\sigma^2 N_c}{2\pi B} \sim 2\,\mbox{GeV}^2\quad\mbox{at}
\quad N_c=3\,,
$$
 if correct, would mean that the resonances essentially overlap (and, 
thus, smear
the spectral density), starting from the first or second excitation. 

\section{How Does All This Match Experiment?}

In spite of the current  extremely high demand on theoretical estimates of
duality violations, there is surprisingly little effort
to elucidate the issue by direct experimental studies.
A high-precision measurement of the ratio $R$ in $e^+e^-$
in a broad range of energies, from threshold  up to, say, 
$s=10\,\, \mbox{GeV}^2$ in a dedicated experiment 
with the proper absolute
normalization would give an enormous boost to this 
issue.  Alas, such measurements have never been undertaken...
The most accurate data on the spectral densities were obtained in
$\tau$ decays \cite{aleph}, where they (naturally) extend only up to
$s=3\,\, \mbox{GeV}^2$. This energy range is way too narrow
to be helpful in perfecting the models which are currently in use
or in the design of new models. 
Still, confronting the data with current theoretic ideas
might give a feeling of whether
or not  we are moving in the right direction.
The comparison is presented in Fig. 11. I will first explain what
 is depicted, and then offer several comments. 

\begin{figure}   
\epsfxsize=11cm
\centerline{\epsfbox{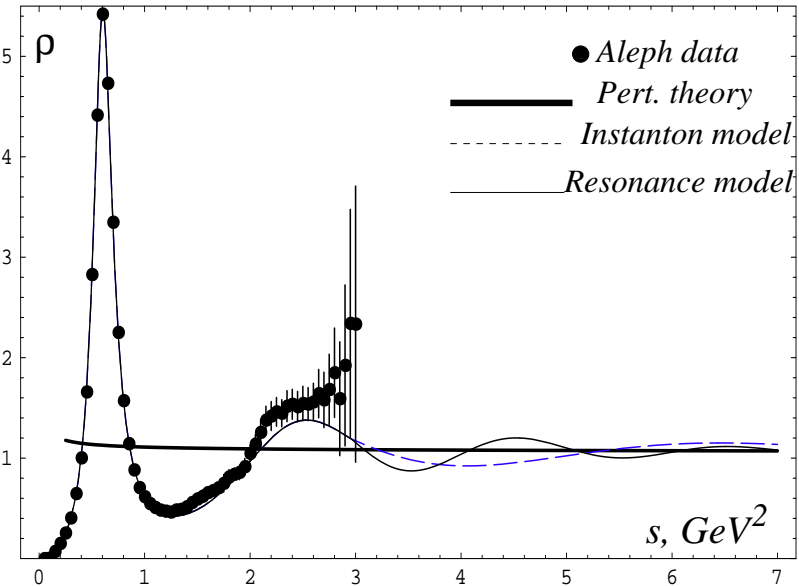}}
 \caption{Experimental data (ALEPH)
on the spectral density in the vector channel and the duality violation models.}
\end{figure}

The ALEPH experimental data \cite{aleph} I show correspond
to  the spectral density in the vector
isovector  channel. 
 There exist  some data above 3 GeV$^2$ from  $e^+e^-$
annihilation, but the   error bars are so large
that plotting these points would just obscure the picture.
Directly measurable in the hadronic $\tau$ decays is the sum of the 
vector and axial spectral densities. To obtain the spectral densities 
separately one has to sort out all decays by assigning
specific quantum numbers to  each given final hadronic state.
In the majority of cases such an assignment is unambiguous.
For instance, two pions (whose contribution is the largest)
can be produced only by the vector current. Some processes, 
however, can occur in both channels, for instance, the $KK\pi$ 
production. Using certain theoretical arguments it was decided
\cite{aleph} that around 3/4 of all $KK\pi$ yield
must be ascribed to the vector channel. Other theoretical arguments 
\cite{EKSV},
which seem  more convincing,
tell that virtually all $KK\pi$ production  takes place in the 
axial  channel. 
Therefore, in fitting the data, I  subtract
the $KK\pi$ yield from the data points presented in Fig. 11. 
I hasten to add that  
the subtraction affects only
the energy range $s>2\,\mbox{GeV}^2$, i.e. the second peak
in Fig. 11, and even in this energy range the effect is small,
$\lsim 5\%$. 
The  subtraction is not essential for a general picture I 
draw here.

The solid curve at  $s <2.7\,\mbox{GeV}^2$ is the best fit
of the data points thus obtained  by the sum of two Breit-Wigner
 peaks (the first one, the $\rho$ meson, is a 
modified Breit-Wigner taking into account threshold effects 
important for the $\rho$ meson). For my purposes
the two-resonance fit presents an excellent approximation
to the experimental spectral density at $s <2.7\,\mbox{GeV}^2$.

The tails of the curves at $2.7< s <7\,\mbox{GeV}^2$ represent the
resonance-based and  instanton-based models (the solid and
dashed curves, respectively)
which I discussed in Secs. 10 and 6.  The solid curve is a modulation of
$\rho(s)_{\rm pert}$ by the factor
\begin{eqnarray}
& & 1 + 1.22\, E^{-3}\,\sin (2\rho E -\delta )\,, 
\nonumber\\[0.3cm]
& &\rho = 3\,\mbox{GeV}^{-1}\,,
\quad \delta = 1.32\,,\quad  E\,\,\,\mbox{in GeV}\,, \nonumber
\end{eqnarray}
cf. Eq. (\ref{homeone}).  The dashed curve is a modulation of
$\rho(s)_{\rm pert}$ by\footnote{As we know from Sec. 10,
the equidistant resonances with
{\em  equal} residues result in oscillations of this type
superimposed on $\rho =1$. By a trivial adjustment of the residues
one can readily achieve that the oscillations (\ref{monone})
are superimposed on $\rho(s)_{\rm pert}$.}  
\begin{eqnarray}
& & 1 - 1.24\, \exp\left( {-\frac{2\pi s B}{\sigma^2 N_c}}
\right)\, \sin \left(\frac{2\pi s }{\sigma^2 } - 3.08
\right)\,,
\nonumber\\[0.3cm]
& &\sigma^2 = 2\,\mbox{GeV}^2\,,\quad B=0.5\,,\quad N_c=3\,,\nonumber
\end{eqnarray}
cf. Eq. (\ref{monone}). 

Finally, the thick solid curve in Fig. 11 displays  $\rho(s)_{\rm pert}$
calculated through order $\alpha_s^3$, with 
$\Lambda^{(3)}_{\overline{\mbox{MS}}}
=  200$ MeV.

It is clearly seen that, qualitatively,  both models match the data well.
In fact, in actuality I would expect an oscillating component
which is a
combination of the solid and dashed curves, since the mechanisms 
of the duality
violations they represent
 are complementary. The instanton mechanism leads to 
a visibly slower (power)
fall off, with rearer oscillations. It is also clear 
that experimental measurements
at the percent level of accuracy will most certainly
provide us with the material needed
for   construction
of a reliable and well calibrated model of the duality
violations.

\section{``Exclusive"  Duality}

This section is somewhat 
perpendicular to the main theme of the review, and can 
be safely omitted. While
previously my topic was estimating duality violations, now I will
discuss a special  ``exclusive" mode of 
{\it implementation}
of the quark-hadron duality. 

In all problems considered above 
 duality is applicable at high energies (momentum transfers),
where the cross sections (decay probabilities)
 are saturated by a large number of 
exclusive channels. Moving in the opposite direction, towards
lower energies (momentum transfers), we decrease
the number of the open channels and, typically,
worsen the accuracy of the quark-hadron duality.
There is a stereotype in the public eye that ``duality works
well in inclusive processes after summing up over a large number
of exclusive channels."
Although this is often true, sometimes the
  stereotype turns out to be wrong. In this section, 
concluding my review, I give two examples when
duality (i.e.  ``equality-with-controlled-accuracy"  
between the quark and hadronic  processes) takes place even though
the hadronic process is saturated by a {\em single} exclusive channel.
Sure enough, such situations
are rather exceptional and are explained by certain
custodian symmetries.

 The first example of this type was discovered
\cite{Shifman:1988rj}
 in the mid-1980's in the heavy quark transitions 
at zero recoil. A prototype process has the form
\beq
Q\to Q'  + \ell\nu
\label{wone}
\eeq 
or, at the hadron level,
\beq
H_{Q} \to H_{Q'}  + \ell\nu\,.
\label{wtwo}
\eeq
Here $Q$ and $Q'$ are heavy quarks of distinct flavors, 
$H_Q$ stands for a $Q$-containing heavy hadron
(practically, the lowest-lying state, say $H_Q= B$ or $H_Q=\Lambda_b$), 
  while
$\ell\nu $ is the lepton pair. 
Needless to say that the sum over all possible $H_{Q'}$'s in the final
state
 is implied since we consider the inclusive reaction.
As we will see shortly,
at zero recoil, only one state will survive
in this sum.

At the point of  zero recoil, the total spatial momentum 
 of the
lepton pair vanishes, $\vec k =0$.
 Then the time component
of the lepton pair momentum in the physical decay
(\ref{wtwo}) must be equal to
$k_0 = M_{H_Q} - M_{H_{Q'}}$. The value of $k_0 $ is maximal
if $H_{Q'}$ is the ground state. It is convenient to introduce
the notation
\beq
\Delta M =\left(  M_{H_Q} - M_{H_{Q'}}\right)_{\rm ground \,\,\, state}\,,
\eeq
and 
\beq
\epsilon = \Delta M - k_0\,.
\eeq
Then on the physical cut $\epsilon$ is real and positive. 

Dynamical information is encoded in the differential probabilities
which go under the name of the hadronic structure
functions $w_i$ (they were already mentioned in the
very end of Sec. 7).
One starts from the transition operator
 describing the forward scattering of $H_Q$ to 
$H_Q$ via intermediate states $H_{Q'}$,
\beq
\hat T_{\mu\nu}  = i\int d^4 x {\rm e}^{-ikx} T
\{\bar Q (x) \gamma_\mu\gamma_5 Q' (x)\, , \,
\bar Q'(0)\gamma_\mu\gamma_5 Q(0)\}\,.
\label{transqcd}
\eeq
For definiteness I limit myself here to the
axial-vector
current, $\bar Q\gamma_\mu\gamma_5 Q'$. 
(The vector current can be treated in a similar fashion.)
   The  hadronic amplitude  is obtained 
by averaging  ${\hat T}_{\mu\nu}$ over  $H_Q$,
\begin{equation}
h_{\mu\nu} =\frac{1}{2M_{H_Q}} \langle H_Q|{\hat T}_{\mu\nu}|H_Q\rangle\,. 
\label{htdef}
\end{equation}
From kinematics
we infer a general decomposition
\begin{equation}
h_{\mu\nu} = -h_1g_{\mu\nu}
+h_2v_\mu v_\nu -ih_3\epsilon_{\mu\nu\alpha\beta}v_\alpha
k_\beta
+h_4 k_\mu k_\nu + h_5 (k_\mu v_\nu + k_\nu v_\mu )\,,
\label{five}
\end{equation}
where $v_\mu$ is the four-velocity of $H_Q$ (in the rest frame
$v_\mu =\{1,0,0,0\}$.) At zero recoil 
$h_i$'s depend only on $k_0$. The physically observable quantities
are
the structure functions 
$$
w_i =2\,{\rm Im}\, h_i \,,\qquad 0\leq\epsilon\leq \Delta M\, .
$$

\vspace{0.1cm}

The relation between $h_i$'s and $w_i$'s is the same as that between
$\Pi (Q^2 )$ and $\rho (s)$ in $e^+e^-$ annihilation. Much in the same way
as in the latter problem,
one can calculate $h_i$ far away from the physical cut,
then perform analytic continuation 
onto real positive $\epsilon$, and then take the imaginary part,
term by term in OPE. A new element  in the heavy flavor decay
is that the expansion runs now in two parameters,
$\Lambda/ \epsilon$ and  $\Lambda/ m_Q$, where $m_Q$
is the heavy quark mass. It is assumed that
$\Lambda /m_{Q,Q'} \ll 1$. Shortly I will
send the quark masses $m_{Q,Q'}$ to infinity
keeping the difference $m_{Q} - m_{Q'}$ fixed. 

\begin{figure}   
\epsfxsize=11cm
\centerline{\epsfbox{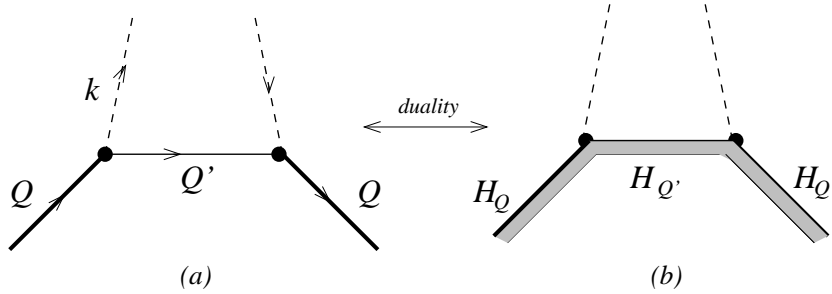}}
 \caption{Inclusive semileptonic decay
of one heavy flavor into another at zero recoil. (a)
Free quark transition operator; (b) hadronic saturation.
The closed circle denotes the insertion
of the axial or vector current. The four-momentum of the lepton pair
is denoted by $k_\mu$.}
\end{figure}

At the quark level $h_i$'s are determined by the diagram displayed 
in Fig. 12(a). 
By analogy with $e^+e^-$ annihilation one might
think that, to make OPE meaningful,
it is necessary to require $\Lambda/ \epsilon \ll 1$.
Surprisingly, at zero recoil this hasty conclusion is wrong.
In fact, it is not difficult to prove (see e.g. \cite{Bigi:1994re})
that the structure of the power expansion in the case at hand
is as follows
\begin{eqnarray}
-h_1 &=& \frac{1}{m_Q-m_{Q'} - k_0}\left(1+ \frac{\Lambda^2}{m_Q^2}+...
\right)\nonumber\\[0.2cm]
&+&\frac{\Lambda}{\left(m_Q-m_{Q'} -
k_0\right)^2}\left(\frac{\Lambda^2}{m_Q^2}+...\right)\nonumber\\[0.2cm]
&+&\frac{\Lambda^2}{\left(m_Q-m_{Q'} -
k_0\right)^3}\left(\frac{\Lambda^2}{m_Q^2}+...\right) + ...\,.
\end{eqnarray} 
If one takes the limit $m_Q\to\infty$ first,
keeping finite the quark mass difference,
\beq
\Delta m \equiv m_Q - m_{Q'} =\mbox{fixed}\,,
\eeq
then the expansion for $h_1$ becomes absolutely  trivial, namely
\beq
-h_1  =  \frac{1}{m_Q-m_{Q'} - k_0} 
\label{fridayone}
\eeq
exactly,  for all values of $m_Q-m_{Q'} - k_0$, large or small.
Thus, at zero recoil, the free quark amplitude of Fig. 12(a) is
exact, a manifestation of the  heavy quark
symmetry. 

Let us now examine $h_1$ from the hadronic side (Fig. 12(b)).
The very same heavy quark symmetry tells
us that the {\em only} state surviving  in the
sum over $H_{Q'}$ 
at $\vec{k}=0,\,\,\, m_{Q'}=\infty$
is the ground state $(Q'\bar q)$ where $q$ stands for the
light spectator quark. Moreover,
\beq
-(h_1)_{\rm hadr}  =  \frac{F^2}{\epsilon}\, ,
\label{fridaytwo}
\eeq
where $F$ is the $H_Q\to H_{Q'}$
form factor at zero recoil. Since $
\Delta M = \Delta m$,
 both expressions (\ref{fridayone}) and (\ref{fridaytwo})
perfectly match provided the transition form factor
is unity at zero recoil \cite{Shifman:1988rj}.
The quark-hadron duality is perfect: the quark pole
of Fig. 12(a) is exactly equal to 
the hadronic pole presented in  Fig. 12(b). Correspondingly, the
same equality takes place for the structure
functions $w_i(k_0)$.

The second example of ``exclusive" duality 
I would like to mention is somewhat more exotic. 
It was found \cite{Bigi:1999qe} as a byproduct
in the studies of the Pauli interference
in the heavy flavor decays in the 't Hooft model.
The setting has  already been presented in Sec. 9,
which is
devoted to the semileptonic decays. Now I pass to
nonleptonic decays. Correspondingly, the weak Lagrangian in Eq.
(\ref{120}) is replaced by
\beq
{\cal L}_{\rm weak} = -\frac{G}{\sqrt{2}}\left\{ a_1\left(\bar c\gamma_\mu b
\right) \left(\bar d\gamma^\mu u
\right) 
+a_2   \left(\bar d\gamma_\mu b
\right) \left(\bar c\gamma^\mu u
\right)  \right\}+\,\mbox{H.c.},
\label{sunone}
\eeq
where I use the same notation as in Sec. 9; two distinct
four-fermion operators in Eq. (\ref{sunone}), with (dimensionless)
 coefficients
$a_1$ and $a_2$, represent two possible patterns of the color flow,
 direct and twisted (in actual QCD the latter emerges due to
hard gluon exchanges). At the quark level
the Pauli interference in the $B^-$ decays
is described by the graph of Fig. 13, which determines
the coefficient in front of the four-fermion operator $\left(\bar
b\gamma_\mu\gamma^5 u
\right) \left(\bar u\gamma^\mu \gamma^5 b
\right) $  in OPE. 
To calculate the effect of the Pauli interference
in the quark language one has to calculate the diagram, take the
imaginary part and
average the four-fermion operator over $B^-$,
\beq
\frac{1}{2M_B}\langle B^-\left|\left(\bar
b\gamma_\mu\gamma^5 u
\right) \left(\bar u\gamma^\mu \gamma^5 b
\right) \right| B^-\rangle =\frac{1}{2}f_B^2M_B\,.
\label{suntwo}
\eeq
If $m_b\gg\beta$, the result thus
obtained has the same status as the parton-model formula, say,  for $R(e^+e^-)$. 

\begin{figure}   
\epsfxsize=7cm
\centerline{\epsfbox{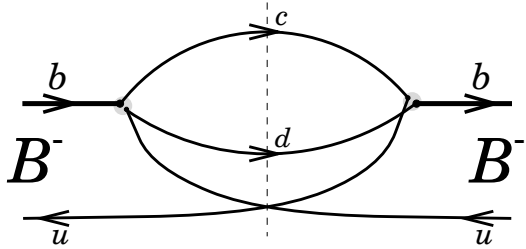}}
 \caption{The quark graph determining
the effect of the Pauli interference in
the nonleptonic inclusive $B^-$ decays at leading order in OPE.}
\end{figure}

Following Ref. 9, I will set all quark masses except that of 
$b$ to zero,
 $m_c=m_u=m_d=0$. Then the Pauli interference
of Fig. 13 takes especially simple form
\beq
\Delta\Gamma_{\rm PI} (B^-) = -\frac{1}{2} (a_1 a_2) G^2\, f_B^2M_B\,.
\label{sunthree}
\eeq
This effect is a $1/m_b$ correction to the total inclusive $B^-$ 
width.\footnote{In actual QCD the Pauli interference would be a $1/m_b^3$
correction;
$1/m_b$ versus $1/m_b^3$ reflects the fact that
the canonic dimension of the current $\bar q\gamma_\mu q$
is one in $D=2$  and three in $D=4$.} At the same time, 
$\Delta\Gamma_{\rm PI}$
is of the same order in $1/N_c$ as $\Gamma_{\rm tot} (B^-)$.
This is important.

As far as the 
hadronic saturation is concerned, the following statements are valid
to   the   leading order in
$1/N_c$: 

\vspace{0.2cm}

 (i) The intermediate states saturating $\Delta\Gamma_{\rm PI}$
are, by necessity, two-meson states. Examples are shown in Fig. 14.

(ii) The graph of Fig. 14(a) is 
due to the first operator in Eq. (\ref{sunone})
and is proportional to $a_1$, while that of
Fig. 14(b) is due to the second operator
and is  proportional to $a_2$. 

(iii) Factorization applies.

\vspace{0.2cm}

Generally speaking, both mesons
in the intermediate state, $\pi^-$ and $D^0$, need not be 
ground states. Radial
excitations would be acceptable,  too. 
However, if  $m_c=m_u=m_d=0$, the
only hadronic states
produced by the currents $ \bar d\gamma^\mu u$ 
and $\bar u\gamma^\mu c$ 
are  the ground  state pseudoscalar mesons with zero mass. 
This is a special feature
of the two-dimensional theory, with no parallel in four dimensions.
The fact that $(d\bar{u})$ meson in Fig. 14(a) is produced by the
$ \bar d\gamma^\mu u$ current forces it to be a massless $\pi^-$,
while, by the same token, the $(u\bar{c})$ 
 meson in Fig. 14(b) is a massless $D^0$.

\begin{figure}   
\epsfxsize=10cm
\centerline{\epsfbox{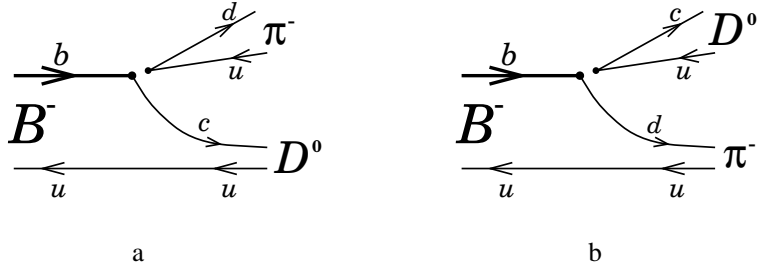}}
 \caption{The product of these two amplitudes determines
$\Delta\Gamma_{\rm PI}$ to the leading order in $1/N_c$.}
\end{figure}

As a result, the only intermediate hadronic state
surviving in the Pauli interference is
\beq
B^-\to \mbox{massless }\,\, \pi^- \,+ \, \mbox{massless }\,\,D^0\to B^-
\eeq
This single exclusive channel   
{\em must} saturate (\ref{sunthree}).

So it does! This comes out rather trivially \cite{Bigi:1999qe}
because of the following properties of the pion coupling and the
$B\to D (\pi )$ transition form factors at $k^2=0$
\begin{eqnarray}
& &f_\pi =\sqrt{\frac{N_c}{\pi}}\int_0^1\phi_\pi (x) \, dx \, =
\sqrt{\frac{N_c}{\pi}}\,,
\nonumber\\[0.2cm]
& &k_\mu \varepsilon^{\mu\nu}
\frac{1}{2M_B}\langle \pi^- |\bar d\gamma_\nu b|B^-\rangle
= -k_z \int_0^1\phi_B (x) \, dx = -k_z\, f_B\, \sqrt{\frac{\pi}{N_c}}\,.
\label{sunfour}
\end{eqnarray}
Here $\phi (x)$ is the light-cone wave function;
for the pion $\phi (x) =1$.
Note the occurrence of $f_B$ in the transition form factor.

Using Eq. (\ref{sunfour}) it is easy to verify that
the massless quark loop of Fig. 14(a) perfectly matches the massless
meson loop, the only contribution to
$\Delta\Gamma_{\rm PI}$ surviving at the hadronic level.
The origin of the factor $f_B^2$ in Eq. (\ref{sunthree})
in the hadronic calculation
is totally different, though; it comes from the square of the transition form
factor, the matching seems magic, yet it could not have happened otherwise.

It is worth stressing  
that the magic simplifications which led to the
``exclusive" mode of the duality implementation
in the problem at hand are explained by miraculous
features of two dimensions.

\section{Conclusions}

In this review I identified
general mechanisms causing deviations from duality
and derived scaling laws governing the
damping of the
duality violating component in a variety of inclusive processes
 as  a
function of energy (momentum transfer). The estimates 
of the absolute normalization one can perform at the moment
 are less certain.
This explains why I was so cautious in my discussion of
the numerical situation in $R_\tau$,
and completely avoided this issue in other inclusive processes.
Getting a better idea of the absolute normalization
is absolutely necessary for all practical applications
of the theoretical constructions presented here.
Precision measurements of $\rho( s)$ in a wide energy range
would be of enormous help in this   question and would,  
probably lead to a speedy solution. 

As I have already noted, duality violations parametrize 
our ignorance. Were an analytic solution of QCD  found,
the contents of this review would become instantly obsolete.
Then the exact asymptotic behavior of 
inclusive cross sections would be known,
and the very concept of duality violations would 
become irrelevant.

\vspace{1cm}

{\bf Acknowledgments}

 \vspace{0.2cm}

\noindent
I would like to thank   N. Uraltsev  and A. Vainshtein  for
useful discussions.

This work was supported in part by DOE under the grant number
DE-FG02-94ER408.

\end{document}